# Rayleigh-Bénard Convection

## Patterns, Chaos, Spatiotemporal Chaos and Turbulence

Tatsuo YANAGITA [1]
Department of Applied Physics, Tokyo Institute of Technology
Oh-okayama, Meguro-ku, Tokyo 152, JAPAN

and

Kunihiko KANEKO [2]
Department of Pure and Applied Science, University of Tokyo
Komaba, Meguro-ku, Tokyo 153, JAPAN

## Contents

[1] E-mail address: yanagita@ap.titech.ac.jp
[2] E-mail address: chaos@tansei.cc.u-tokyo.ac.jp

# List of Figures

## Abstract

A coupled map lattice for convection is proposed, which consists of Eulerian and Lagrangian procedures. Simulations of the model not only reproduce a wide-range of phenomena in Rayleigh-Bénard convection experiments but also lead to several prediction of novel phenomena there: For small aspect ratios, the formation of convective rolls, their oscillation, many routes to chaos, and chaotic itinerancy are found, with the increase of the Rayleigh number. For large aspect ratios, the collective oscillation of convective rolls, travelling waves, coherent chaos, and spatiotemporal intermittency are observed. At high Rayleigh numbers, the transition from soft to hard turbulence is confirmed, as is characterized by the change of the temperature distribution from Gaussian to exponential. Roll formation in three-dimensional convection is also simulated, and found to reproduce experiments well.

# 1 Introduction

Rayleigh-Bénard convection has been extensively studied as a "standard" experimental model for temporally and/or spatially complex phenomena. When the aspect ratio is small, it shows a variety of routes to chaos (subharmonic, quasi-periodic and intermittent), depending on the Prandtl number. For large aspect ratios, spatiotemporal intermittency is observed, which provides one of the standard routes from localized to spatiotemporal chaos, as is common in spatially extended systems. When the Rayleigh number is very large, the experiments provide a test-bed for turbulence theory. It includes the recent discovery of the transition from soft to hard turbulence, found by Libchaber's group. In cases with a large aspect ratio and a relatively low Rayleigh number, pattern formation of rolls has been extensively studied.

In principle, it is expected that these experiments can be described by the Navier-Stokes (NS) equation, coupled suitably to an equation for the temperature field. In a weakly nonlinear regime, for example, a set of equations comprising of the NS equation and a temperature field with Boussinesq approximation is in quantitative agreement with experimental observations, with regards to the critical Rayleigh number and the onset of oscillation of rolls. In a highly nonlinear regime (chaotic and turbulent regime), one has to resort to numerical simulation, to study this set of equations. However, it is often difficult to use these equations there, because of the limitation of computational resources and numerical stability problems, practically. Furthermore, it is sometimes not easy to understand the phenomenology for convection completely, even if we succeed in reproducing the phenomena.

So far, we do not have a "simple" model which reproduces all of the above phenomena. Is it possible, then, to construct a simpler (and coarse grained) model to study the phenomena?

In this paper, we introduce a coupled map lattice model which reproduces almost all phenomena known for Rayleigh-Bénard convection (see [1] for the rapid communication of the present paper). With this model we can study the salient features of convection interactively, and discuss the mechanisms of a variety of transition sequences of convection patterns. Also, we can analyze the phenomena, in terms of dynamical systems, with the use of, for example, Lyapunov analysis. Another advantage of this model is its fast computation. All the simulations are carried out with the use of workstations, rather than a CRAY or Connection Machine, although our model fits with parallel computations very well. This numerical efficiency enables us to globally search a wide parameter space, to predict a new class of phenomena, and even to make some quantitative predictions.

The present paper is organized as follows: In §2, we construct a CML model for convection by introducing the Lagrangian procedure which expresses the advection for the flow. Numerical results of the model are presented through §3 to §9. The onset of convection (Rayleigh-Bénard instability) and the onset of periodic oscillations is studied in §3. With the increase of the Rayleigh number, the periodic oscillations are replaced by chaotic ones. In § 4, a variety of routes to chaotic oscillations is found, in agreement with experiments. It is also argued that the interruption of period-doubling bifurcations, experimentally observed, may not be due to external noise, but inherent to the dynamics of convection, which originally involve many degrees of freedom. After these low dimensional attractors collapse with the increase of the Rayleigh number, chaotic itinerant motion between the collapsed attractors is often observed, as is studied in § 5.

For large aspect ratios, spatial degrees of freedom are no longer suppressed. High-dimensional chaos is observed whose dimension increases with the system size. We note that the spatial structure is still sustained here, leading to coherent chaos. With the increase of the Rayleigh number, a transition to a state with spatial disorder is seen universally. This route to turbulence, characterized by spatiotemporal intermittency (STI), is confirmed in our model in § 7, with a detailed statistical analysis. The transition from soft to hard turbulence is studied in § 8. Experimental observations regarding the change of distributions are reproduced, while a mechanism for the transition is proposed with the visualization of plumes. A prediction on the Prandtl-number dependence of the transition is also given. The pattern formation process in convective rolls is given in § 9, as well as the inclusion of a rotation effect. A summary and discussions are given in § 10. In appendix A, we discuss the stability of our model, by showing that the salient feature does not depend on the detailed procedure of our model.

# 2 Model

Coupled map lattices (CML) are useful for studying the dynamics of spatially extended systems [2][3][4][5]. Originally, the CML was proposed as a model for studying spatiotemporal chaos at a rather abstract and metaphorical level. However, the results derived from this model are often strongly connected with those found in natural phenomena. For example, spatiotemporal intermittency (STI) was first found in a class of CML, and is observed in a wide range of CML models when the system loses spatial coherence and moves towards turbulence. Later STI was also discovered in systems with partial differential equations (PDE) such as the Kuramoto-Sivashinsky equation and the Ginzburg-Landau equation. In nature, such STI behavior is observed, e.g, in Rayleigh-Bénard convection, electric convection of liquid crystals and rotating viscous fluids.

As far as we see from the examples of STI, the qualitative features do not depend on the details of the models. Some other features, found in abstract CML models are also found widely in PDE systems and in experiments. These observations lead us to believe in the existence of qualitative universality classes in nature. Without bothering about the details of the equations involved, we may construct a simple model for some given spatially extended dynamics. Here we provide an example of the construction of a simple model which potentially belongs to the same "universality class" as Rayleigh-Bénard convection.

CML modeling is based on the separation and successive operation of procedures, which are represented as maps acting on a field variable on a lattice. Besides the above mentioned abstract case, this approach has successfully been applied to spinodal decomposition [6], the boiling transition [7][8], pattern formation of sand ripples [9], and so on. In particular, the pattern formation process derived from a CML model of spinodal decomposition has been shown to form a universality class including the time dependent Ginzburg-Landau equation, and agrees with experimental observations, even in a quantitative sense with regard to scaling relationships.

Let us start with the construction of a CML model for convection in 2-dimensional space. For this, first we choose a two dimensional lattice $(x, y)$ with $y$ as a perpendicular direction, and assign the velocity field $\vec{v}^t(x, y)$ and the internal energy $E^t(x, y)$ as a field variables at time $t$. The dynamics of these field variables consists of Lagrangian and Eulerian parts. The latter part is further decomposed into the buoyancy force, heat diffusion and viscosity, which are carried out by the conventional CML modeling method [10]. In constructing procedures, we assume that $E^t(x, y)$ is associated with the temperature.

## 2.1 Euler procedure

To construct the procedures in the Eulerian part we take into account of the following properties in convection phenomena; (1) Heat diffusion leads to diffusion of $E^t(x, y)$; (2) The velocity field $v^t(x, y)$ is also subjected to the diffusive dynamics, due to the viscosity; (3) A site with higher temperature receives a force in the upward direction (buoyancy); (4) The gradient of a pressure term (which depends on the velocity field) gives rise to a change of the velocity fields.

The procedures for (1) and (2) are rather transparent, since we can just adopt the discrete

Laplacian procedure of diffusively coupled map lattices. The construction of (3) and (4) is more subtle and difficult. For the buoyance procedure, we assume that the vertical velocity is incremented linearly with the horizontal Laplacian of the energy term. Indeed we have tried some other procedures also, such as a Laplacian term also including the vertical direction. In so far as we have studied, our choice here fits best with known results on the convection (see Appendix A).

To take (4) into account, we note that the pressure term requires $div\vec{v}$ to be 0, in an incompressible fluid. We do not use this condition here, since the inclusion of pressure variables requires more complicated modeling, and often makes it difficult to construct a model with local interaction only. Instead, we borrow a term from compressible fluid dynamics, which brings about this pressure effect, and refrains from the growth of $div\vec{v}$. This term is given by the discrete version of $grad(div\vec{v})$.

Combining these dynamics, the Eulerian part is written as the successive operations of the following mappings (hereafter we use the notation for discrete Laplacian operator: $\Delta A(x,y) = \frac{1}{4}\{A(x-1,y)+A(x+1,y)+A(x,y-1)+A(x,y+1)-4A(x,y)\}$ for any field variable $A$):

(a) Buoyancy procedure

$$v_y^*(x,y) = v_y^t(x,y) + \frac{c}{2}\{2E^t(x,y) - E^t(x+1,y) - E^t(x-1,y)\} \tag{1}$$

$$v_x^*(x,y) = v_x^t(x,y) \tag{2}$$

(b) Heat diffusion

$$E'(x,y) = E^t(x,y) + \lambda\Delta E^t(x,y) \tag{3}$$

(c) Viscosity and pressure effect

$$v_x'(x,y) = v_x^*(x,y) + \nu\Delta v_x^*(x,y) + \eta\{(v_x^*(x+1,y)+v_x^*(x-1,y))/2 - v_x^*(x,y) + (v_y^*(x+1,y+1)+v_y^*(x-1,y-1)-v_y^*(x-1,y+1)-v_y^*(x+1,y-1))/4\} \tag{4}$$

and the equation with $[x \leftrightarrow y]$

The above three parallel procedures completes the Eulerian scheme.

## 2.2 Lagrange procedure

The Lagrangian scheme expresses the advection of velocity and temperature. To take advection into account, it is often useful to introduce a quasi-particle on each lattice site $(x, y)$. The particle has a velocity $\vec{v}(x, y)$ and moves to $(x + \delta x, y + \delta y)$ by the Lagrangian scheme, where $\delta x = v_x(x, y), \delta y = v_y(x, y)$. All field variables (velocity and internal energy) are carried by this particle. Since there is no lattice point at the position $(x + \delta x, y + \delta y)$ generally, we allocate the field variable on its four nearest neighbor sites. The weight of this allocation is given by the lever rule; $(1 - \delta x)(1 - \delta y)$, $\delta x(1 - \delta y)$, $(1 - \delta x)\delta y$, and $\delta x \delta y$ for the sites $([x + \delta x], [y + \delta y]), ([x + \delta x] + 1, [y + \delta y]),([x + \delta x], [y + \delta y] + 1)$ and $([x + \delta x] + 1, [y + \delta y] + 1)$ respectively, with $[z]$ the largest integer smaller than $z$ (see figure 1 for the explanation). By this rule, the energy and momentum are conserved in the Lagrangian procedure.

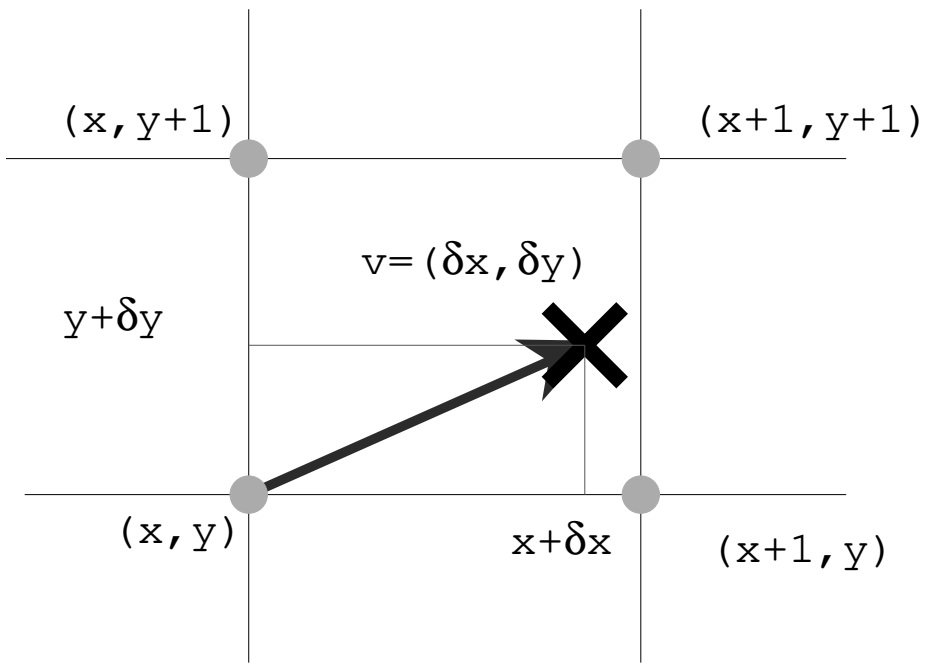

Figure 1: Lagrangian procedure

Schematic figure illustrating the Lagrangian procedure. A quasi-particle sets at each site $(x, y)$ moves to $(x, y) + (v_x, v_y)$, following the velocity field at the original site. Then the field values are allocated to the nearest neighbor's sites, according to the lever rule.

The total dynamics of our model is given by successive applications of the above procedures;

$$\left\{ \begin{array}{c} \vec{v}^t(x, y) \\ E^t(x, y) \end{array} \right\} \xmapsto{(a)} \left\{ \begin{array}{c} \vec{v} * (x, y) \\ E^t(x, y) \end{array} \right\} \xmapsto{(b),(c)} \left\{ \begin{array}{c} \vec{v}'(x, y) \\ E'(x, y) \end{array} \right\} \xmapsto{\text{Lagrangian}} \left\{ \begin{array}{c} \vec{v}^{t+1}(x, y) \\ E^{t+1}(x, y) \end{array} \right\}$$

This completes one step of the dynamics.

For the boundary, we choose the following conditions;

(1) Top and Bottom plates: Assuming a correspondence between $E$ and the temperature, we choose the boundary condition $E(x, 0) = \Delta T = -E(x, N_y)$. For the velocity field we have used either the fixed boundary or free boundary. For the Lagrangian scheme, we use either the fixed or the reflection boundary.

(2) Sidewalls at $x = 0$ and $x = N_x$: We use either fixed, reflective, or periodic boundary conditions. Hereafter we mostly choose the fixed boundary for top and bottom plates and the periodic boundary condition for the $x$-direction. The change to a fixed boundary at the wall alters our velocity pattern at least in the small size case, but most of the transition sequences of patterns (to be reported) remain invariant.

The basic parameters in our model (and in experiments) are the Rayleigh number (proportional to $\Delta T$), the Prandtl number (ratio of viscosity to heat diffusion $\nu/\lambda$), and the aspect ratio ($N_x/N_y$). Here we study in detail the dependence of convection patterns on the Rayleigh number and the aspect ratio, and mention the effect of Prandtl-number for each transition. For simulations we take the diffusion coefficient $\eta$ as $\nu/3 < \eta \leq \nu$ [11], although our results are reproduced, as long as $\eta$ is in the same order of magnitude as $\nu$, where $div\vec{v}$ is kept small numerically.

## 2.3 Defense of our CML approach

One might ask why we do not use a set of PDE with the Navier-Stokes (NS) equation and a suitable heat equation for the temperature field. There are several reasons for this. First, one has to resort to numerical simulations to solve the NS equation, since it cannot be treated analytically at least in a high Rayleigh number regime. The numerical scheme for solving it is also complicated and often it is unstable. To stabilize the numerical scheme, we often have to add artificial viscosity, for example; a higher order term than the Laplacian. Without such a method, we cannot avoid the numerical viscosity which is due to the discreteness of our computation. Such an "artificial viscosity" drastically changes the functional analytical property of the NS equation [12][13]. For example, the NS equation does not have a physically or mathematically unique solution, while inclusion of the higher order viscosity leads to a unique global solution. So it is not clear whether the numerical solution (even if it is carried out with high accuracy) retains the mathematical structures of the NS equation. On the other hand, it is not completely sure whether the NS equation is *the* equation for fluid dynamics [13, 14]. As is clear from the above argument, numerical solutions in agreement with experiments do not justify the NS equation, while one may argue that the functional analytical properties of the NS equation may not fit our physical intuitions. Recall that the NS equation is not derived from a microscopic level with complete exactness. It is a phenomenological equation, like the Ginzburg-Landau equation in pattern formation dynamics[14].

In Rayleigh-Bénard convection, we also have to adopt some approximation for the coupling with the temperature field. Thus the PDE equations there remain approximate and phenomenological.

The CML model we adopt here is constructive in nature. For this construction we assume that the salient features of the phenomena do not depend on the details of a model. A model, at any rate, cannot be exactly identical with nature herself, and we have to assume some kind of universality among the model classes. Most important macroscopic (coarse grained) properties such as the flow patterns and statistical quantities must be robust against some changes of models. Thus we can hope that our CML modelling belongs to the same "universality" class as convection in nature. Conversely, by modifying or removing the procedures in our model, we can see what parts are essential to a given feature. Numerical results of other models (with modification and removal of some procedures) are given in Appendix A, where the predominance of the present model is discussed, as well as the stability against the change of models.

Since our model is much more efficient than the PDE approach, it is easy to explore the phenomenology globally with changing parameters. A Rayleigh-Bénard system has at least three basic parameters. We do not know as yet how long it takes to explore all the phenomenology in convection with the use of the PDE approach, even if we use the fastest computer in the world. In our model, we can easily study the phenomena interactively with work-stations, by exploring the three-dimensional parameter space. Surprisingly, the model reproduces almost all phenomenology for convection as is shown in the following sections. Furthermore we can even get some predictions for future experiments. In particular, we can predict some features of the turbulence regimes, which are rather difficult to study by PDE simulations, so far.

Of course, another merit of our approach is its easy accessibility of dynamical systems theory. For all classes of convection patterns to be studied, considerations are made from the point of dynamical systems theory, by which we can proceed to the understanding of convection patterns.

# 3 Pre-Chaos

## 3.1 The onset of Convection

When $\Delta T$ is sufficiently small, there is no convective motion; the fluid is completely fixed in time, while the Fourier law of the temperature is numerically confirmed. When $\Delta T$ is increased, the heat transfer by the diffusion is no longer enough to sustain the temperature difference, and convective rolls start to appear (figure 2). The critical temperature difference $\Delta T_c$ for the appearance of convection corresponds to the critical Rayleigh number in experiments. The critical value depends on the boundary condition at small aspect ratios.

Slightly above $\Delta T_c$, the convective rolls are fixed in time. In this subsection, we investigate several features at the onset of convection.

In figure 3 the vertical velocity in the middle of the container $v_y(N_x/2, N_y/2)$ is plotted vs $\Delta T$. At $\Delta T = \Delta T_c$ the vertical velocity rises from zero. To see the critical property here, we have to determine $\Delta T_c$ numerically with accuracy. Here we use the following method: Let us measure the time evolution of the total kinetic energy

$$K(t) = \sum_{x=1}^{N_x} \sum_{y=1}^{N_y} \{v_x^t(x,y)^2 + v_y^t(x,y)^2\} \tag{5}$$

starting from a random initial condition. If the total kinetic energy decreases with time then $\Delta T$ is lower than $\Delta T_c$, otherwise $\Delta T > \Delta T_c$. By measuring the time derivative for the total kinetic energy $dK/dt$, the critical value $\Delta T_c$ can be determined by the condition $dK/dt = 0$. By using this critical value, and the following normalized temperature difference (corresponding to the normalized Rayleigh number)

$$\epsilon = \frac{\Delta T - \Delta T_c}{\Delta T_c}, \tag{6}$$

the vertical velocity $v_y$ is found to scale as

$$v_y(\epsilon) \sim \epsilon^{1/2}. \tag{7}$$

This exponent 1/2 is, of course, expected from the Hopf bifurcation analysis, and agrees with experiments [15][16].

When there is convection ($\Delta T > \Delta T_c$), the effective thermal conductivity $\lambda_{eff}$ of the convective layer is greater than the static thermal conductivity $\lambda$. The Nusselt number

$$Nu \equiv \lambda_{eff}/\lambda \tag{8}$$

is 1 for $\epsilon < 1$, and is known to satisfy $Nu - 1 \propto \epsilon$ [16][17]. This relationship of the Nusselt number is confirmed in our simulation, while the proportion coefficient A ( s.t. $Nu-1 = A\epsilon$) depends on the aspect ratio and the condition of the side walls of the container. At the onset of instability, critical slowing down is commonly observed when a constant heat flux is supplied. To study this, we define the heat current as a constant increment (decrement) of energy at the bottom (top) plate of the container. We have measured the time evolution of the temperature difference $\Delta T(t)$ between the top and the bottom plates. When a heat current is turned on at time 0, heat is initially carried only by thermal diffusion. Before $\Delta T(t)$ reaches its maximum value, the temperature difference exceeds the critical value

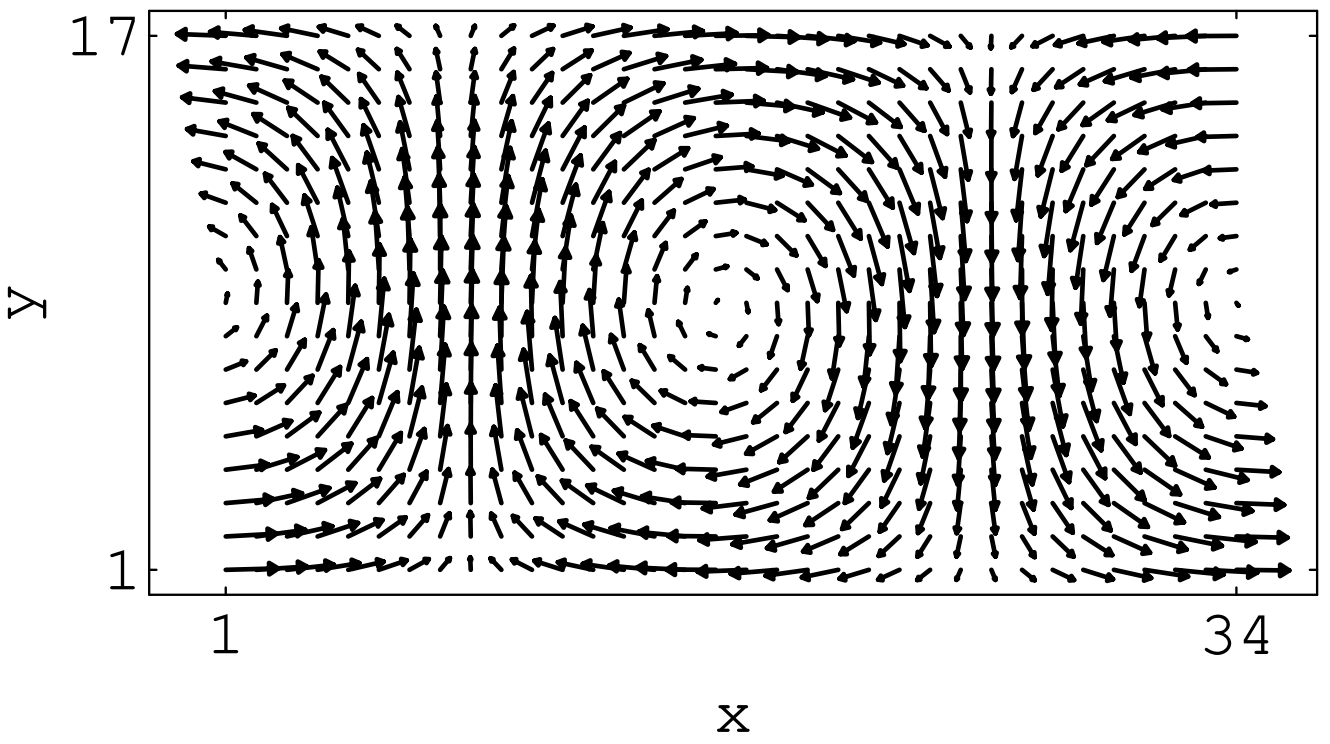

Figure 2: Vector field of convective rolls near the onset of convection

The snapshot of the vector field of convective rolls near the onset of convection. These rolls are fixed in time. $\Delta T = 0.01, \lambda = 0.4, \nu = \eta = 0.2, N_x = 34, N_y = 17$ with periodic boundary conditions after the transients have died out. A random initial condition is adopted.

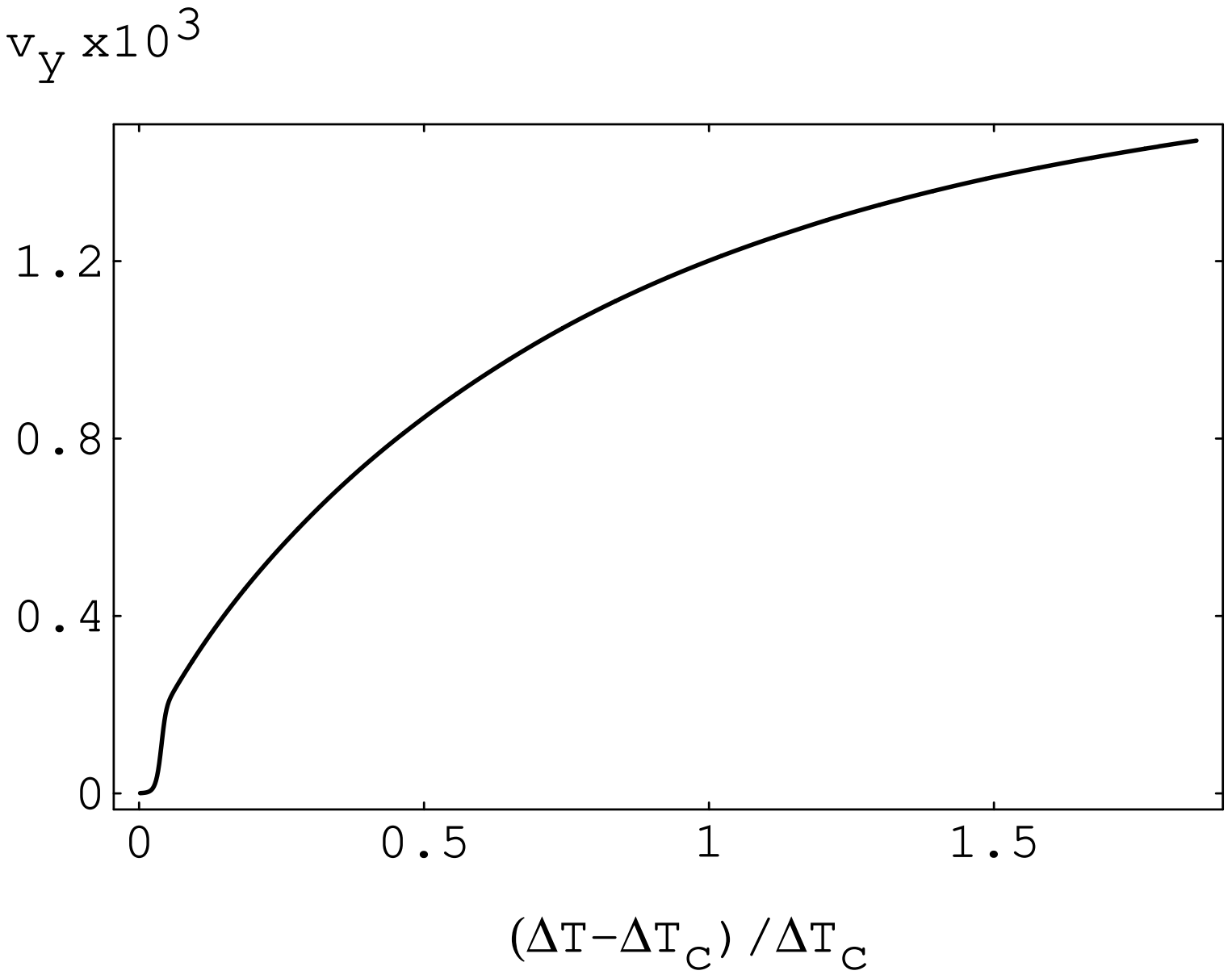

Figure 3: The $\Delta T$ dependence of the vertical velocity

The vertical velocity $v_y(N_x/2, N_y/2)$ in the middle of the container as a function of the normalized temperature difference. $\Delta T$ is gradually incremented form 0.007 to 0.02 per 0.00013. The each $v_y(N_x/2, N_y/2)$ is obtained after 1000 transients. Above $\Delta T_c \sim 0.007$, $v_y$ suddenly increases with some power. $\lambda = 0.4, \nu = \eta = 0.2, N_x = 34, N_y = 17$.

$\Delta T_c$. Then the fluid starts the convection by which the temperature difference starts to decrease. The final decay to equilibrium can be represented by

$$\Delta T(t) = D \exp(-\frac{t}{\tau}) + \Delta T(\infty). \tag{9}$$

We have fitted our data with the above form in order to determine $\tau$. Figure 4 shows the heat flux dependence of the decay time $\tau$. The inverse of the decay time linearly increases with the heat as is expected, and agrees with experiments and theory [16].

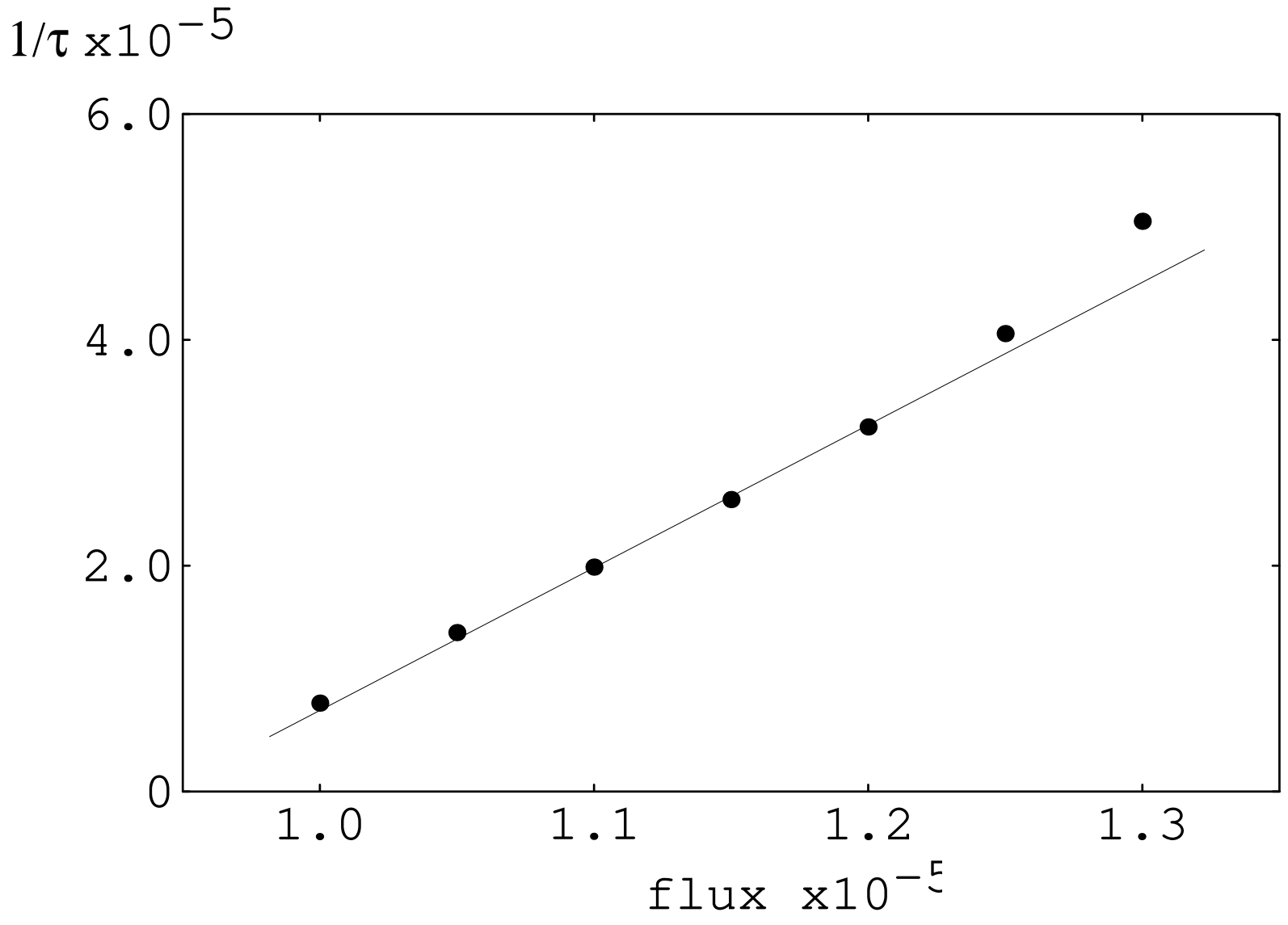

Figure 4: The heat flux dependence of the relaxation time

Applying a constant flux, we calculate $\Delta T(t)$ until $2 \times 10^5$ time steps starting from a random initial condition. We fit the time evolution of the temperature difference $\Delta T(t)$ by equation (9). The inverse of $\tau$ increases with the heat flux linearly which vanishes at the onset of convection. $\lambda = 0.4, \nu = \eta = 0.2, N_x = 34, N_y = 17$

## 3.2 The onset of Oscillations

By further increasing $\Delta T$, the rolls are no longer fixed, but start to oscillate. The amplitude of the oscillation increases in proportion to $\Delta T - \Delta T_{osc}$ near the onset of the oscillation $\Delta T_{osc}$. The critical temperature difference at the onset of the oscillation $\Delta T_{osc}$ depends on $Pr$, $\Gamma$ and the boundary conditions. First, $\Delta T_{osc}$ is proportional to the Prandtl number. Changing $\lambda$ from 0.05 to 0.4 by fixing $\nu = \eta = 0.2, N_x = 34, N_y = 17$, the critical value of the onset of oscillation $\Delta T_{osc}$ is fitted as;

$$\Delta T_{osc}(\lambda) = 1.8\lambda + 0.3.$$

The aspect ratio dependence is, on the other hand, nonmonotonic. When the aspect ratio $\Gamma$ equals an integer, the rolls start to oscillate at a small temperature difference. If $\Gamma$ is not an integer, motion of the convective rolls should be restricted by the unmatched size of the container.

At the onset, the oscillation is periodic without any higher harmonics of the fundamental frequency. The characteristic frequency of the oscillation also depends on $\Delta T$, $Pr$ and the aspect ratio $\Gamma$. We have studied the dependence of the frequency on $\Delta T$, near the onset of the oscillation. To estimate the power spectrum of the velocity, the AutoRegressive model (AR model) is adopted (see Appendix B)[18].

From the AR model one can estimate the oscillation frequency as well as the amplitude of the fundamental frequency by the difference between the maximum and minimum vertical velocities $v_y(N_x/2, N_y/2)$. The amplitude $A_{max}$ is found to be proportional to $\Delta T$ near the onset of oscillation (figure 5). By increasing $\Delta T$, the maximum peak of the power spectrum (fundamental frequency) shifts to a higher frequency. Figure 6 shows the $\Delta T$ dependence of the characteristic frequency. The fundamental frequency almost linearly increases with $\Delta T$.

# 4 Routes to Chaos

For small aspect ratios, the periodic oscillation may bifurcate to chaos via several routes as the Rayleigh number is changed. Such routes to chaos have been compared with dynamical systems theories, and have been observed in numerous experiments over the past few decades [19][20]. The other system parameters, Prandtl number and the aspect ratio are known to be relevant to the nature of the bifurcation.

At a low Prandtl numbers, we have found a subharmonic route to chaos. With the increase of the Rayleigh number, the period of the oscillation doubles as is shown in figure 7. Here, we note that the doubling is interrupted after a finite number of times (so far the maximum periodicity we observed is 16). In experimental observations, it is believed that noise induces such an imperfect period-doubling bifurcation cascade. In our simulation, no external noise is added, and high dimensional dynamics possibly plays the role of a generator of “noise”, which, we believe, is the origin of the interruption of the doubling sequence.

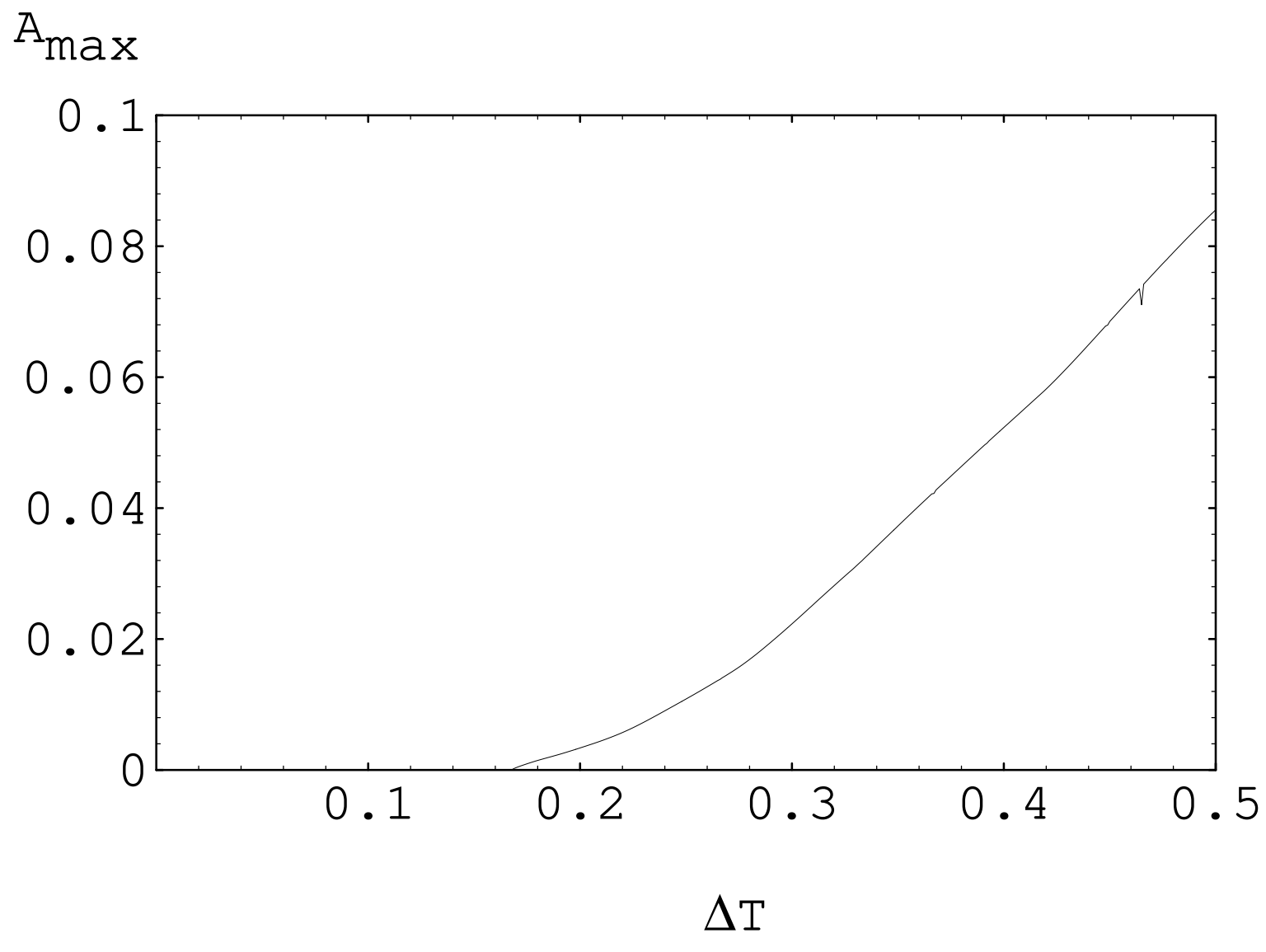

Figure 5: The $\Delta T$ dependence of the amplitude of oscillation

The amplitude of the oscillation vs $\Delta T$ is plotted. The amplitude is defined as the difference between maximum and minimum values of $v_y(N_x/2, N_y/2)$ after discarding $10^4$ initial transients. $\lambda = 0.4, \nu = \eta = 0.2, N_x = 34, N_y = 17$.

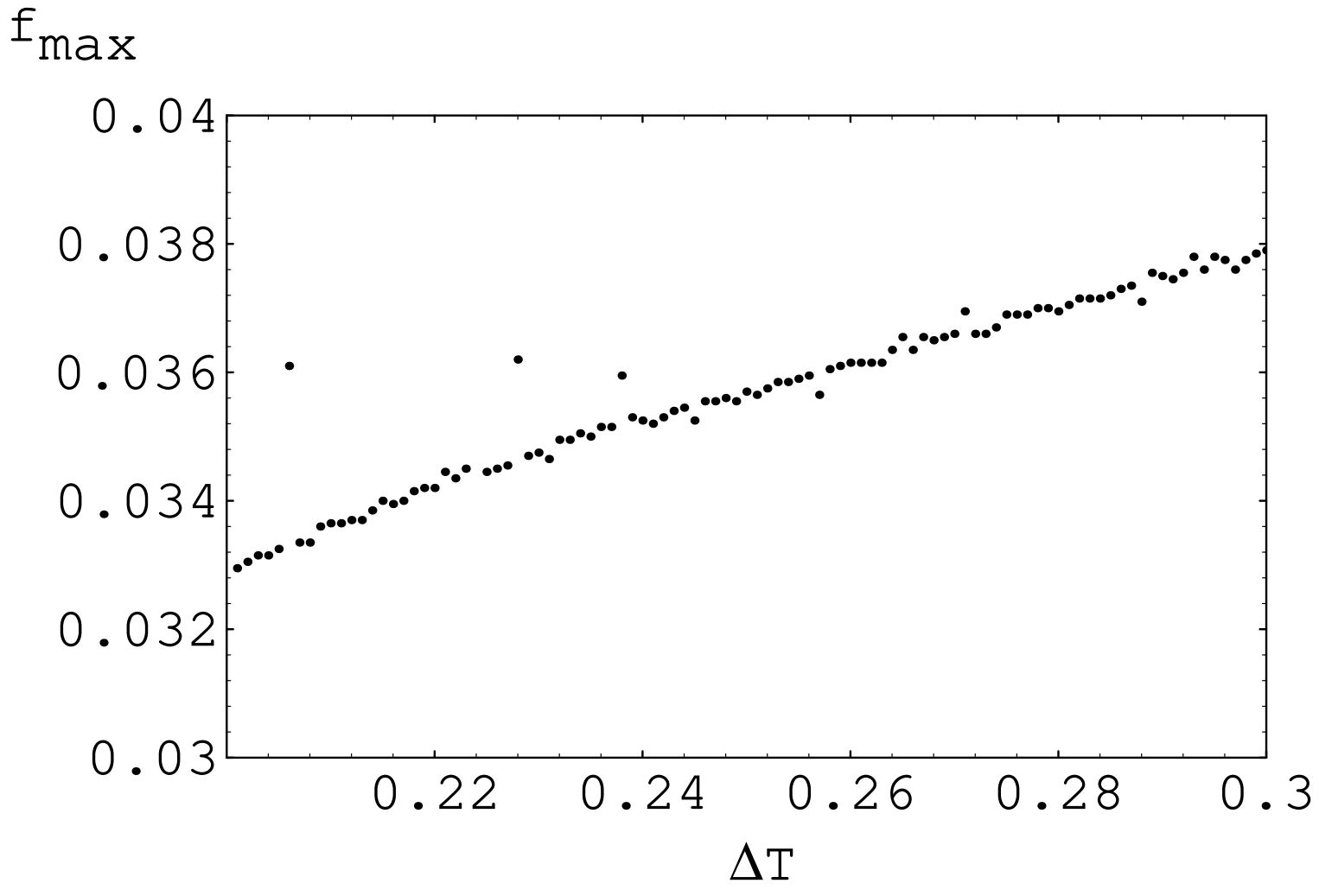

Figure 6: The $\Delta T$ dependency of the oscillation frequency

The fundamental frequency vs $\Delta T$. The each fundamental frequency is estimated by 100th order ARmodel (see Appendix B) using 4000 time series per 10 steps of the velocity $v_y(N_x/2, N_y/2)$ after $10^4$ transients have died out. All the parameters are same as in figure 5.

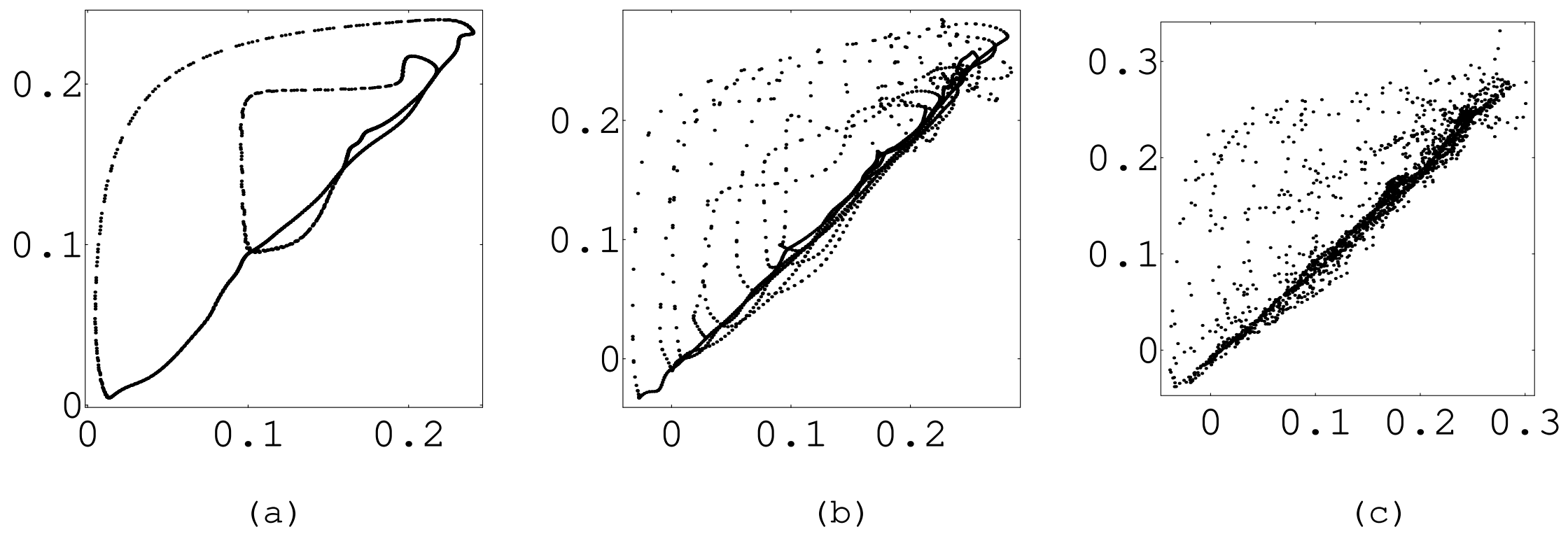

Figure 7: The projection of the orbit of the vertical velocity $v_y(N_x/2, N_y/2)$

The change of the oscillation of convective rolls with the increase of $\Delta T$, from periodic to chaotic. The projection of the orbit the vertical velocity $v_y^t(N_x/2, N_y/2)$ versus $v_y^{t+10}(N_x/2, N_y/2)$. 4000 time series are plotted. (a) $\Delta T = 0.5$:period 2 (b) $\Delta T = 0.55$: period 8 (c) $\Delta T = 0.6$:chaotic. $\lambda = 0.4, \nu = \eta = 0.2, N_x = 30, N_y = 17$.

At high Prandtl numbers, we have found a quasiperiodic route to chaos (see figure 8), as well as the mode locking phenomena near the collapse of tori. At an even higher Prandtl number, torus doubling is often observed [21]. These changes of routes to chaos are consistent with experimental observations [22]. The route to chaos with intermittency is also observed by changing the aspect ratio.

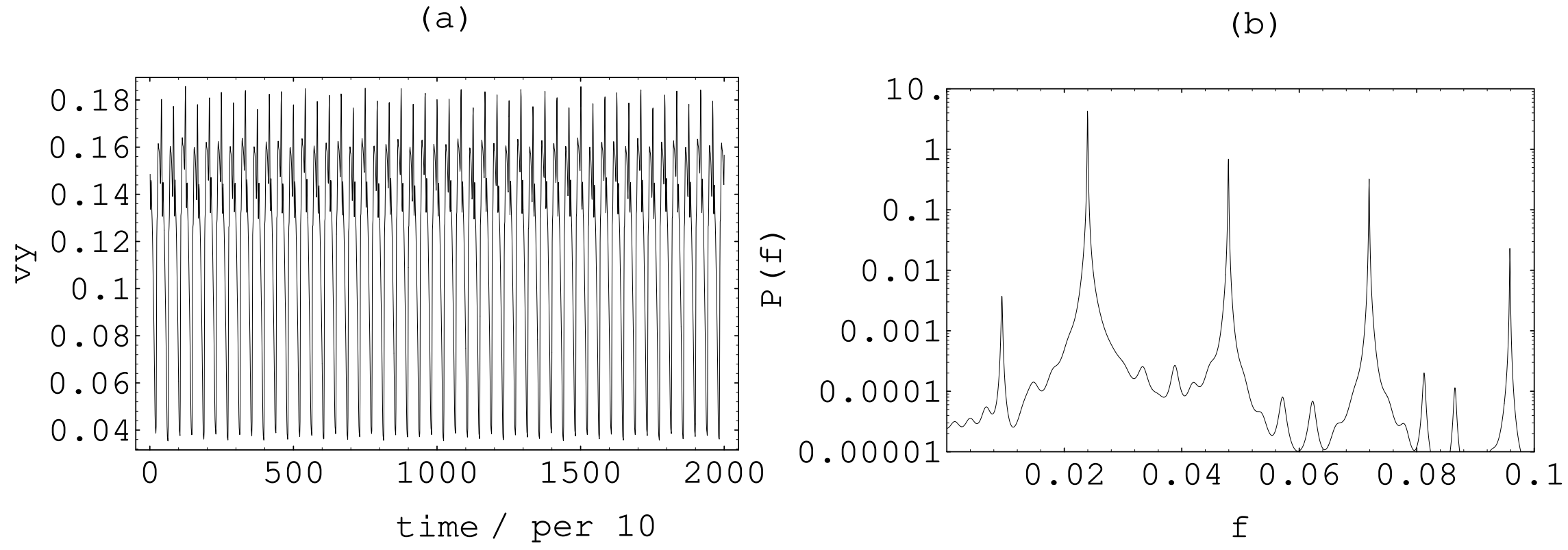

Figure 8: Quasiperiodic route to chaos

(a) At high Prandtl numbers, the motion of convective rolls is quasiperiodic. The time series of the vertical velocity $v_y^t(N_x/2, N_y/2)$ is plotted at $\Delta T = 0.2, \lambda = 0.1, \nu = \eta = 0.2, N_x = 34, N_y = 17$. (b) Power spectrum for the above time series (a). Two elementary frequencies exit.

These routes to chaos strongly depend on the system parameters, in particular, on the aspect ratio $\Gamma$. With the change of $\Gamma$, the number of convective rolls can also vary. For example, with the further increase of $\Delta T$, (after the oscillation of two convective rolls becomes chaotic), we have sometimes observed the periodic oscillation of three rolls. This "window" is due to quite a different mechanism from that in low-dimensional chaotic dynamical systems. In the present case, the change of spatial degrees of freedom (the number of convective rolls) is a trigger for the bifurcation to a periodic state. Furthermore, the "window" here has a hysteresis with respect to the changes of $\Delta T$ and $Pr$. Thus the route to chaos can depend on the history of the variation of the parameters.

The change of the number of rolls is rather abrupt with the change of the aspect ratio. Once the number changes, the low-dimensional dynamics governing the motion alters drastically, which can push the attractor from chaotic to periodic motion. Since the change of roll structures has a hysteresis with $\Delta T$ and Prandtl number, we can observe a different route to chaos for the same parameters, depending on the history.

# 5 Chaotic Itinerancy

In the previous section, we have seen that the routes to low-dimensional chaos in our simulations agree well with experiments. In this section, we investigate how well the correspondence with experiments holds further into the high-dimensional region with spatial structures, and study how "turbulent" motions appear after the low-dimensional "chaotic" behavior. Here we see how the change from low-dimensional to high-dimensional dynamics occurs as a change from low-dimensional chaos to turbulence, where the dimension of the attractor is much higher. In other words, turbulence is regarded to be chaotic not only in time but also in space, and can be called spatiotemporal chaos. We consider how the spatial structure of the convective rolls collapses, especially in a confined system with a relatively small aspect ratio.

Here we study the chaotic change of roll patterns, observed by increasing $\Delta T$ beyond the chaotic regime. With this phenomenon, we see a switching behavior between low-dimensional and high-dimensional motions.

An example of the switching phenomena is given in figure 9, where the rotational direction of the convective rolls switches intermittently in time. Over a long time interval, the convection pattern remains regular, until a disorganized motion in space and time appears. After this "turbulent" motion, the convective pattern comes back to a regular one, while the direction of the flow often is reversed (figure 10).

When $\Delta T$ is slightly lower than that for this switching phenomenon, two attractors exist which correspond to the upward and downward streams of rolls. Depending on the initial conditions one of these attractors is selected (two different basins exist). These separate attractors are connected to form a single attractor when $\Delta T$ exceeds the threshold $\Delta T_{CI}$ for the switching behavior. Beyond $\Delta T_{CI}$, almost laminar spatial structures (corresponding to one of the attractors for $\Delta T < \Delta T_{CI}$) suddenly break down and a turbulent motion (disordered in space) appears. Then either one of the patterns corresponding to the attractors for $\Delta T < \Delta T_{CI}$ is selected and the motion is laminar again. This process of switching continues for ever, as far as we have observed. If the state between two "laminar" regions were described by low-dimensional chaos, this phenomenon would be described as a bifurcation with symmetry breaking (restoration), which is a rather common one. In the present case, the motion between the "laminar" states involves many degrees of freedom, as will be confirmed. Thus the behavior here cannot be described by a low-dimensional

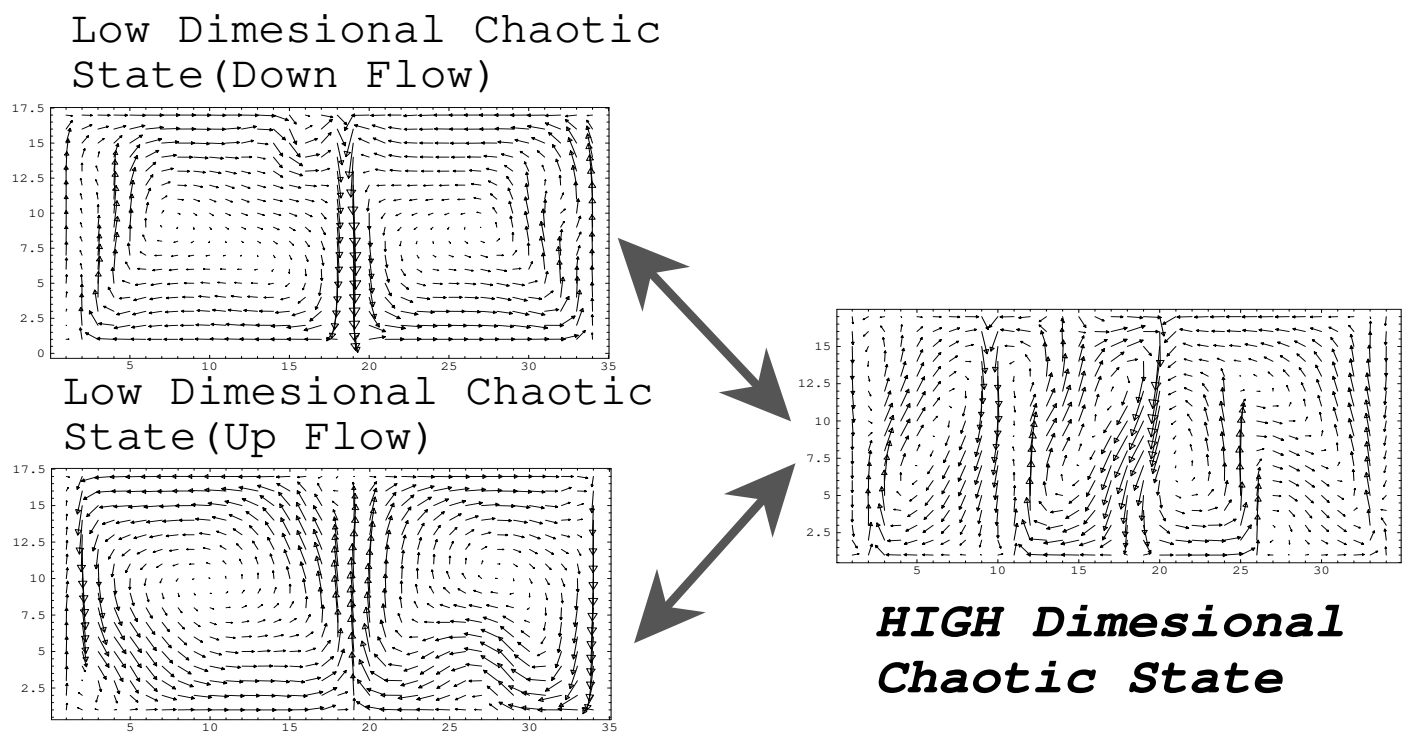

Figure 9: Switching between collapsed attractors

Snapshots of the vector field during chaotic itinerancy. Almost laminar convective rolls suddenly collapse and are replaced by turbulent motions, until a new direction of convective rolls is selected after the turbulence. The switching between upstream and downstream occurs intermittently. $\Delta T = 2.0, \lambda = 0.02, \nu = \eta = 0.2, N_x = 34, N_y = 17$.

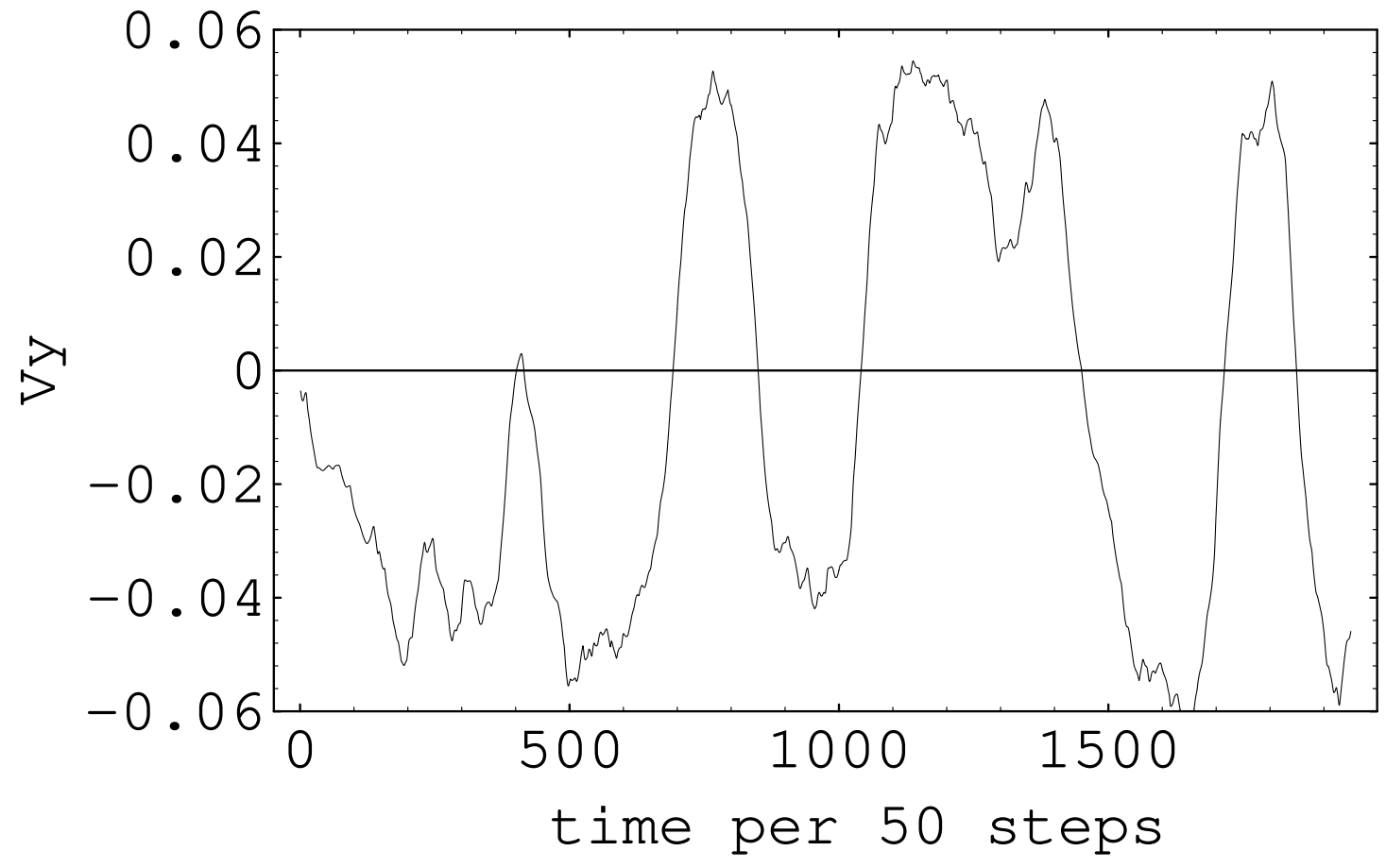

Figure 10: Time series of the vertical velocity in the middle of the container

Temporal evolution of (short-time) average of the vertical velocity $v_y(N_x/2, N_y/2)$ is plotted per 50 steps, which shows intermittent switching between upward and downward directions. Here we take 50 time steps for the average $(v_y^{t-50} + v_y^{t-50+1} + \cdots + v_y^t)/50$. In the course of the switching to a new rotational direction, highly turbulent behavior is observed. $\Delta T = 1.5, \lambda = 0.02, \nu = \eta = 0.2, N_x = 34, N_y = 17$.

dynamical system. Indeed, this type of behavior here has never been observed for an ODE model reduced from the Navier-Stokes equation, with taking only few numbers of Fourier modes.

The itinerancy over low-dimensional ordered motions via high-dimensional turbulent motions is known as *chaotic itinerancy* and has been observed in a variety of dynamical systems, including globally coupled maps [23] [24], Maxwell-Bloch turbulence [25], neural dynamics [26], and also in an optical experiment [27]. Similar phenomena as CI has also been observed and analyzed in global climate dynamics [28].

So far there have been no reports on chaotic itinerancies in Bénard convection. This, we believe, is due to the fact that convection experiments are often focused either on low-dimensional chaos or on a very high-dimensional dynamics. Thus we predict that the behavior here should be observed by studying the intermediate situation.

This chaotic itinerancy motion is studied quantitatively, by using a probability distribution for a life time of laminar and turbulent states. In order to get such a binary representation, first we define the number of rolls which exist in the container. The number of rolls can be estimated by the number of local maxima of the function $f(x) \equiv v_y(x, N_y/2)$. We call the motion turbulent if the number of rolls exceeds a threshold $n_c$, and otherwise call it laminar [3]. The life time distribution of these states exhibits quiet a different type of behavior with the increase of $\Delta T$ (figures 11 and 12).

Both the distributions of the turbulent and the laminar states are exponential. The characteristic life time of the turbulent states is almost independent of $\Delta T$. On the other hand, the life time of the laminar state is much longer and increases by decreasing $\Delta T$ until it diverges at the critical point $\Delta T_{CI}$, where the two laminar states are disconnected. The exponential distribution of the turbulent state implies that the state plays the role of "loss of memory" in the course of this motion. Indeed, the rotational direction is almost randomly selected by losing the memory of the previous laminar state. Furthermore, the time for the selection of the direction of streams (up/down-ward) is almost independent of $\Delta T$.

We have computed the Lyapunov spectrum [29], to characterize the switching, and to estimate the involved degrees of freedom during turbulence. The Lyapunov spectrum measures the averaged divergence of nearby trajectories in phase space. The number of positive Lyapunov exponents gives a rough measure for the effective number of "degrees of freedom".

[3] As long as we take the threshold $3 < n_c < 7$, statistical properties we study do not depend on this choice of $n_c$.

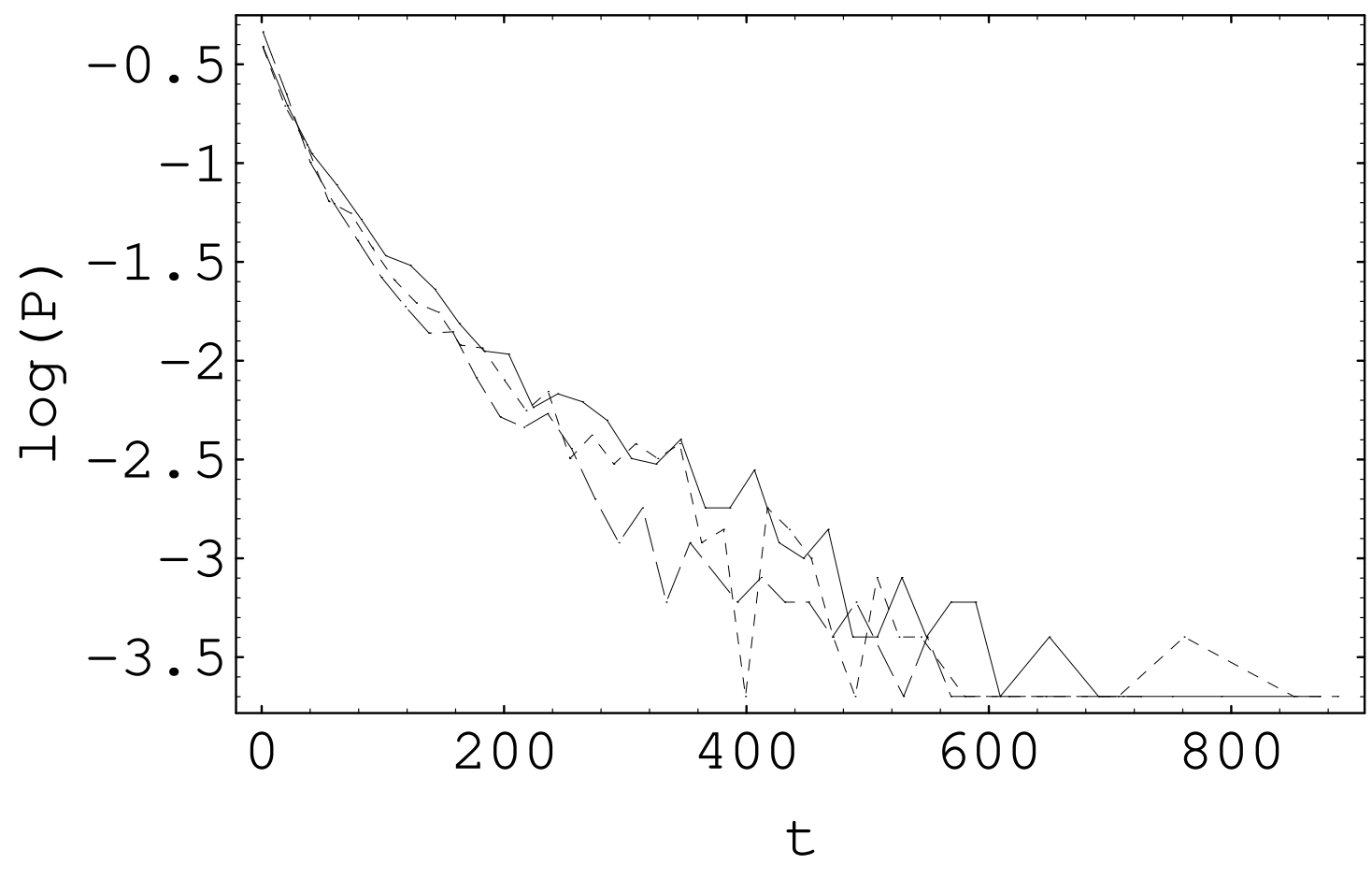

Figure 11: The life time distribution of the turbulent state

Semi-log plot of the life time distribution of the turbulent state. The distribution is taken over $10^4$ residence time. The form of the distribution does not depend on the temperature difference between the top and bottom plates. Solid line:$\Delta T = 0.07$, doted line:$\Delta T = 0.09$, broken line:$\Delta T = 0.15$. The other parameters are same as in figure 10.

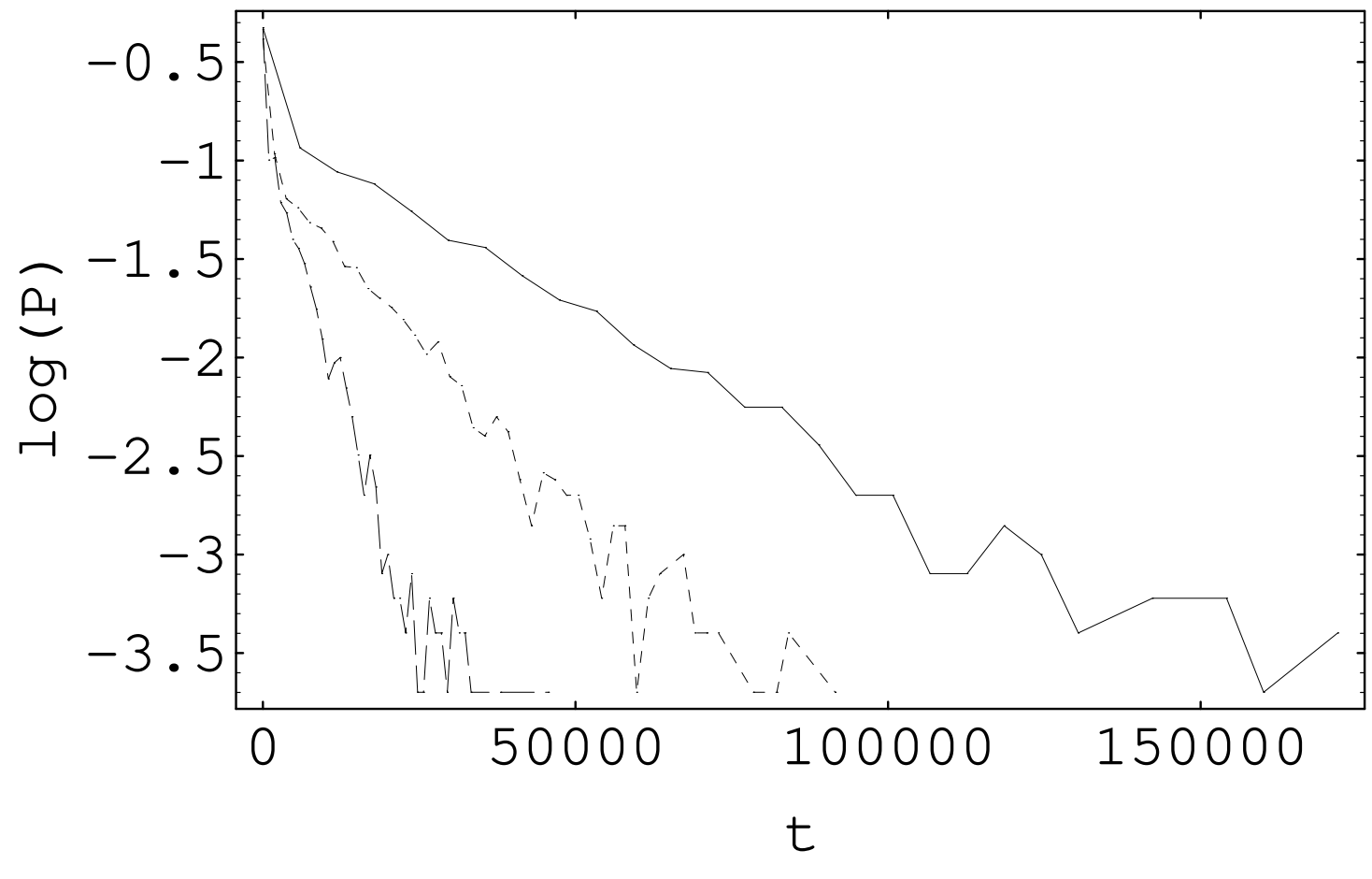

Figure 12: The life time distribution of the laminar state

The distribution of the life time of the laminar state, which should be compared with figure 11. The distribution is taken over $10^4$ residence time. The characteristic life time is clearly dependent on $\Delta T$. Solid line:$\Delta T = 0.07$, doted line:$\Delta T = 0.09$, broken line:$\Delta T = 0.15$. The other parameters are same as in figure 11.

Figure 13 shows the Lyapunov spectra in different $\Delta T$ at the chaotic itinerancy regime. The number of positive Lyapunov exponents is almost constant at around 7, over the range of $\Delta T$ from 0.5 to 2.0. Thus 7 chaotic modes are involved in the motion. By increasing $\Delta T$, the life time of the laminar state gradually decreases, and the dynamics of convective rolls gets complex towards developed spatiotemporal chaos. However, the number of positive Lyapunov exponents is constant here, which means that the number of unstable directions in phase space is almost constant in the chaotic itinerancy region. It may also be useful to point out that the shape of Lyapunov spectra are rather flat at the null exponents. Such trace of the plateau at the null exponent may represent a cascade process (e.g., successive split of vortices) at the collapse of (low-dimensional) ordered motion [30][31].

The Lyapunov spectra may not be useful for distinguishing the chaotic itinerancy from the usual chaotic motion, since they are obtained from long (infinite) time averages. In order to characterize the dynamic properties, we have computed local Kolmogorov-Sinai entropy (LKSE). The Kolmogorov-Sinai entropy (KSE) is estimated by the sum of the positive Lyapunov exponents [32].

$$KSE = \sum_{j=1}^{q} \lambda_j$$
$$(\lambda_q > 0, \lambda_{q+1} \leq 0) \quad (10)$$

Instead of averaging over a long time, we define the local Lyapunov spectrum by the average over a given finite (short) time interval here. By using the local Lyapunov exponents, we define the LKSE as the sum of the local positive Lyapunov exponents. Thus the LKSE is time dependent, and characterizes some dynamic features. Time series of LKSE are plotted in (figure 14), which has a spiky structure. Each peak in this time series corresponds to the turbulent motion of the convective rolls.

From the local Lyapunov exponents, one can get some information on the dynamics of the degrees of freedom. The number of positive local Lyapunov exponents gives a measure for the degrees of freedom at each time. Indeed, the time series of this number shows essentially the same behavior as that of the LKSE. Thus the switch between two low-dimensional states via high-dimensional motion is confirmed.

In the present paper, we have focused on the chaotic itinerancy motion with two convective rolls. However, we have observed the same chaotic itinerancy behavior for the cases with 3 or 4 rolls. Generally this type of behavior is observed at low or intermediate aspect ratios (e.g., $\Gamma < 5$). Here turbulent behavior appears after a few numbers of rolls is selected. At these aspect ratios, spatiotemporal chaos appears through the chaotic itinerancy motion of convective rolls. First, we observe turbulence as spatiotemporal chaos as a switching state between two laminar (but temporally chaotic) states, and then, with the increase of $\Delta T$, the portion of such turbulent motion increases. Since this behavior is rather generally observed, we expect that it will also be observed in experiments, by choosing a suitable aspect ratio and Rayleigh number.

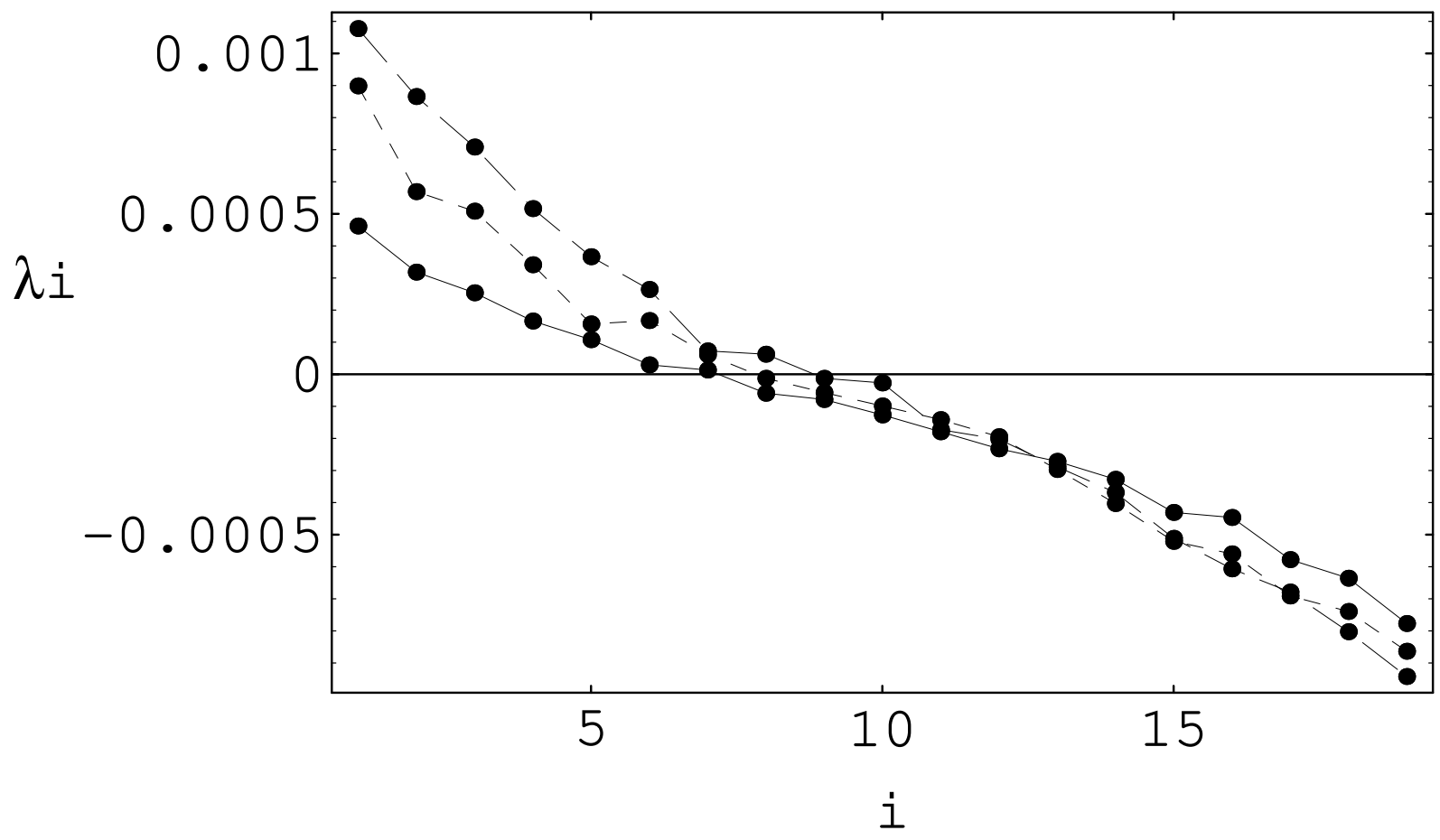

Figure 13: Lyapunov spectra in the chaotic itinerancy region

The first 20 Lyapunov exponents are plotted, computed by the average of $10^5$ time steps. In the chaotic itinerancy regime, the number of positive Lyapunov exponents is almost constant with the increase of $\Delta T$, while the value of the positive exponents increases with $\Delta T$. Solid line:$\Delta T = 0.5$, dotted:1.0, broken:1.5.

# 6 Coherent Chaos

So far we have studied a system with a relatively low aspect ratio, where Rayleigh-Bénard convection has provided a good test system for the study of the transition to chaos, since the spatial degree of freedom for convective rolls is suppressed. Indeed we have confirmed the universal routes to chaos (subharmonic, quasi-periodic and intermittent).

If the aspect ratio is much larger, the number of rolls is very large. Since the spatial degrees of freedom roughly correspond to the number of convective rolls in the Rayleigh-Bénard system, the spatial degrees of freedom are no longer suppressed. Here it is difficult to observe the universal routes to temporal chaos discussed in the previous section. In this case, a study of the transition to "spatiotemporal chaos" is required.

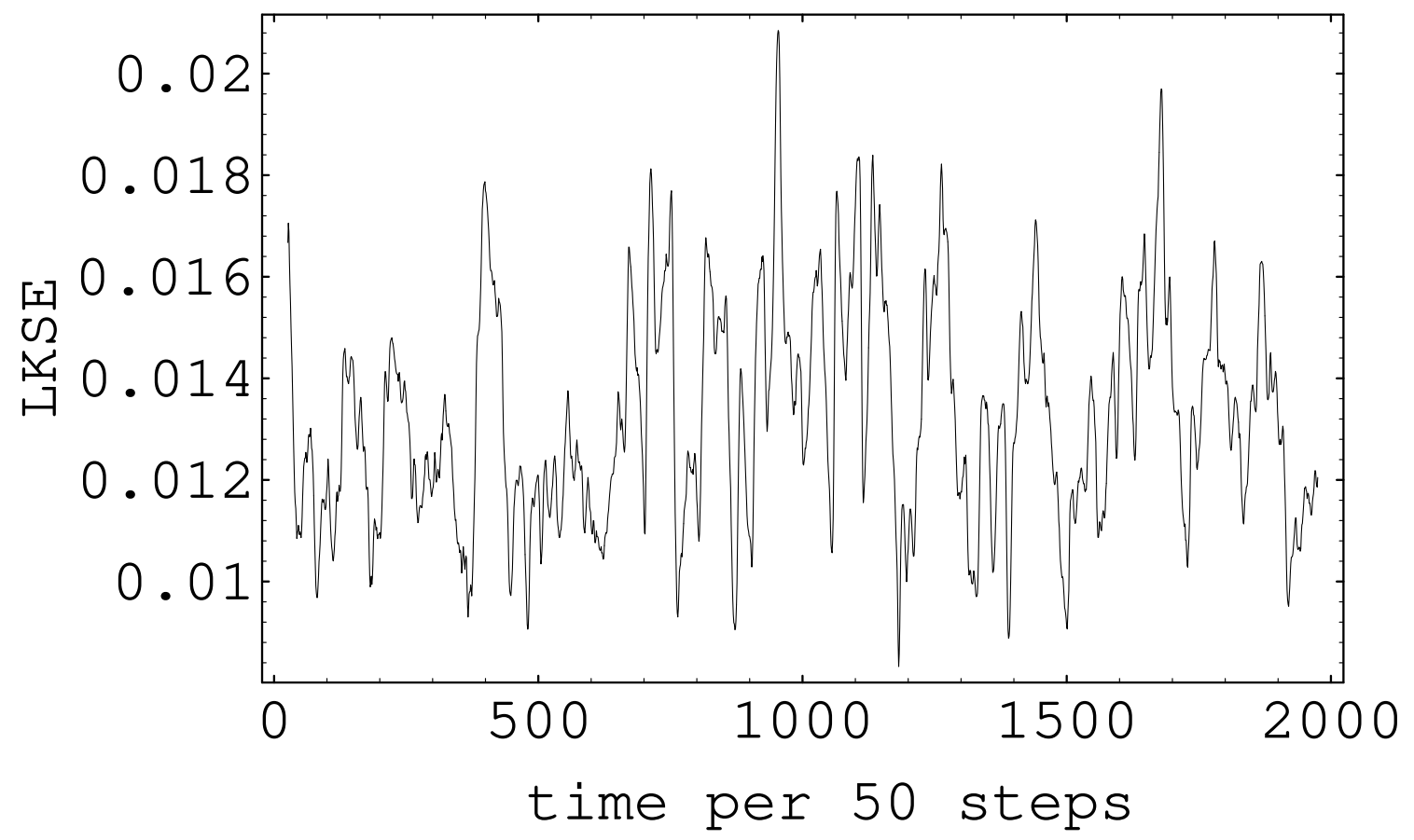

Figure 14: Time series of local KS entropy

Time series of the LKSE estimated by the sum of the local positive Lyapunov exponents. Each LKSE is obtained as the average over 2500 time steps around each time step. For each collapse of the convective rolls, a sudden the increase of the LKSE is observed. $\Delta T = 1.5, \lambda = 0.02, \nu = \eta = 0.2, N_x = 34, N_y = 17$

When $\Delta T$ is increased above the critical value $\Delta T_c$, a perfect chain of convective rolls is formed. The number of rolls increases with $\Delta T$. For example, the number of rolls increases from 10 ($\Delta T = 0.001$) to 13 ($\Delta T = 0.02$) with $\Gamma = 10$ ($N_y = 17$). With the further increase of $\Delta T$, the number of rolls starts to vary in time. During the selection process of the number of rolls, the long time transient behavior (glassy motion of rolls) has been observed.

Above some threshold, these rolls oscillate collectively (all rolls oscillate with almost the same frequency). We have measured the temperature in the middle of the container along the horizontal direction. The local maxima correspond to the positions of hot streams while the local minima to cold streams in the vertical direction. In figure 15, the positions of the local maxima and minima are plotted in spacetime. It can clearly be seen that the positions of hot and cold streams oscillate collectively. In addition, cold streams and hot streams oscillate with opposite phases. The value of the local maximum also oscillates in time. Such a collective oscillation is also observed in experiments [33]. We have not found any aspect-ratio dependence of the oscillatory frequency while changing $\Gamma$ from 10 to 30. Thus the oscillation is governed not globally but locally with interactions of neighboring cells.

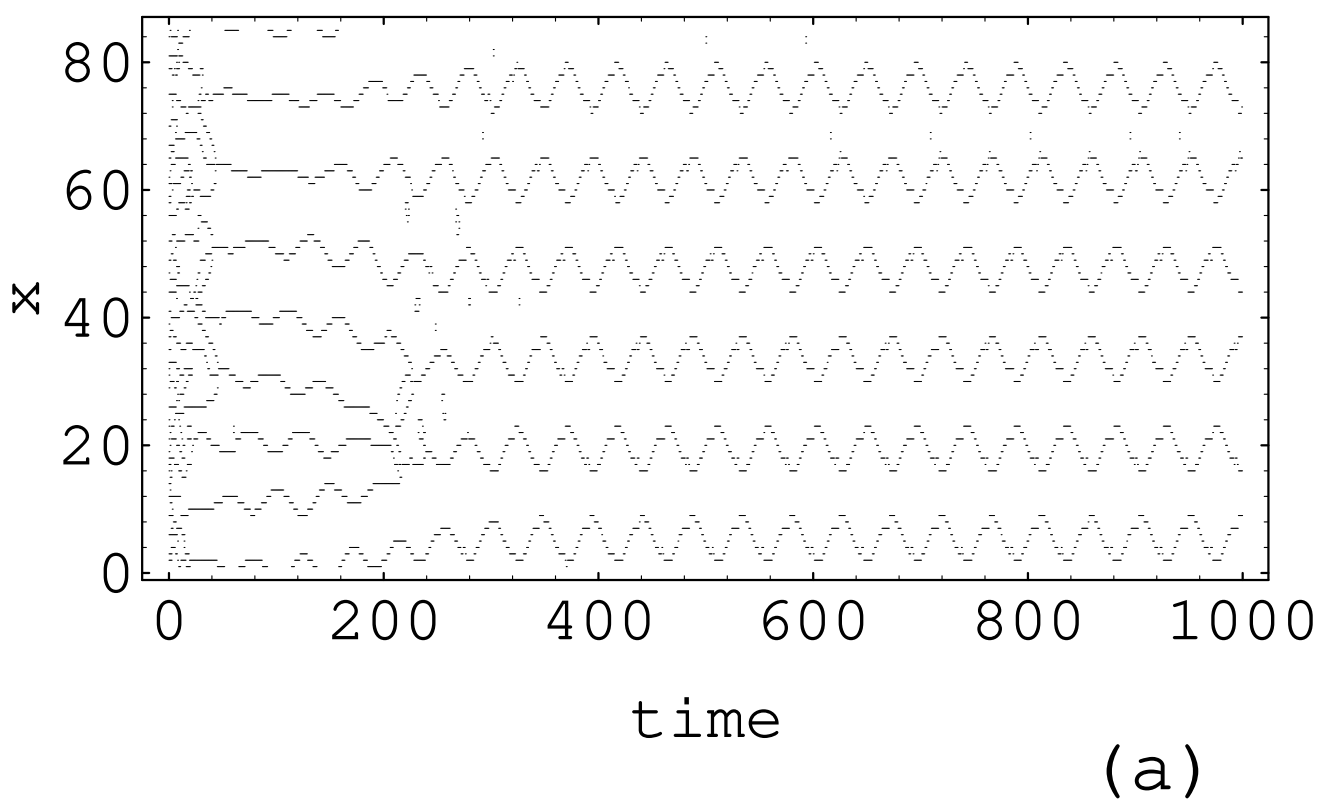

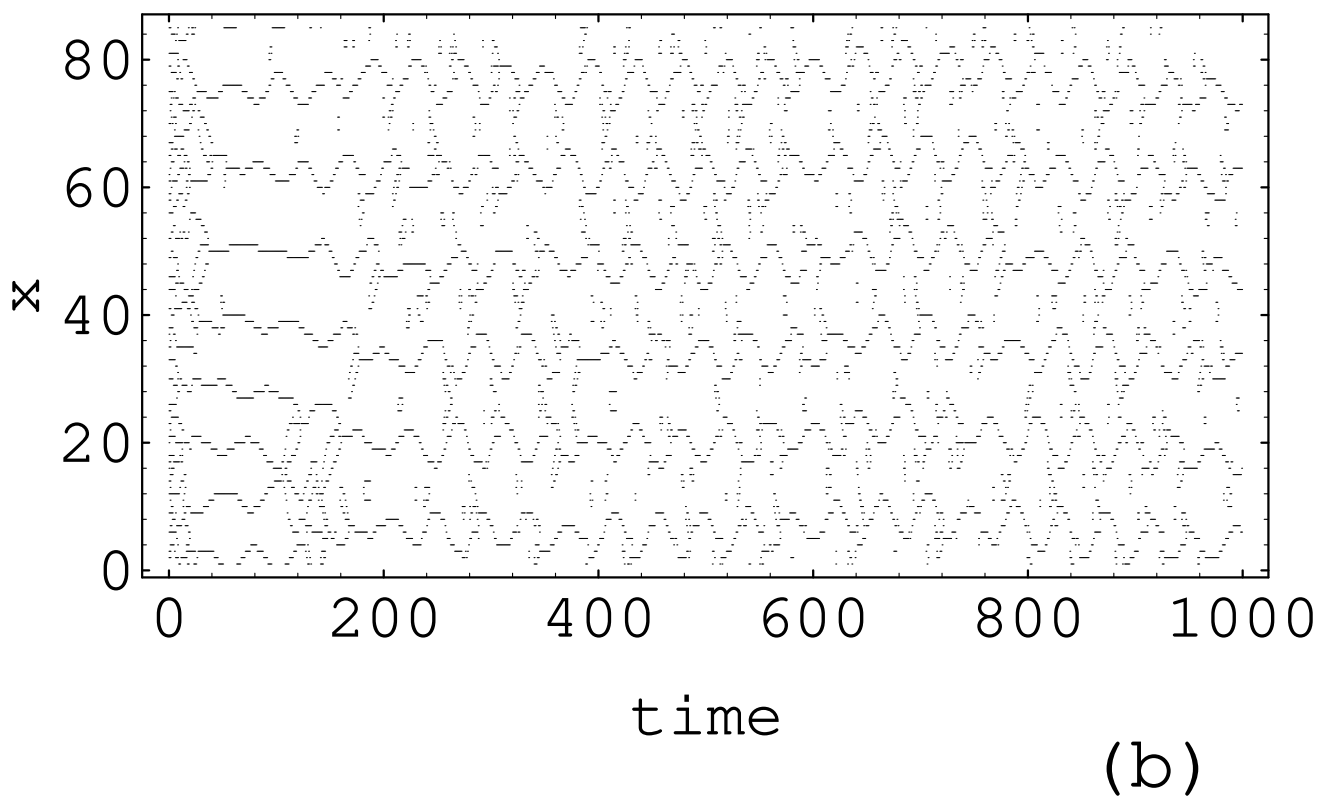

Figure 15: Collective oscillation of convective rolls

Spacetime positions of convective rolls are plotted. At large aspect ratios, collective oscillation of rolls is observed. Cold and hot streams oscillate with opposite phases. As $\Delta T$ is increased, turbulent patches appear in the spacetime diagram, providing STI. $\lambda = 0.02, \nu = \eta = 0.2, N_z = 17, N_y = 85$, (a):$\Delta T = 0.04$, (b):$\Delta T = 0.05$.

We have found chaotic motion with spatial coherence at intermediate values of $\Delta T$ between collective and STI behavior (see next section). We confirm the existence of spatial coherence by the spatial correlation $C(x)$;

$$C(x) = \langle v_y^t(x_0, N_y/2) \cdot v_y^t(x_0 + x, N_y/2)\rangle. \tag{11}$$

In figure 16, the spatial correlation starts to decay up to some distance, beyond which it converges to a finite value (unequal to zero). Thus the spatial coherence is sustained with a regular structure of convective rolls.

On the other hand, the motion is chaotic with many unstable modes. We have measured the Lyapunov spectrum by changing the aspect ratio from 2 to 50. The number of positive Lyapunov exponents increases in proportion to the aspect ratio. The Lyapunov spectrum $\Lambda(x) \equiv \lambda_{\Gamma x}$ scaled with the aspect ratio, as is shown in figure 17, approaches a single form when the size is increased. Such scaling behavior is often seen in spatiotemporal chaos [34][35].

Thus a system with a large aspect ratio consists of a chain of chaotic oscillators, and the dimension is expected to diverge linearly with the system size. Hence one may expect that the spatial coherence (phase relationship between rolls) may be lost. This is not the case. The coherence is maintained as can be seen is $C(x)$. The reason for this "coherent" chaos is due to the separation of scales. Here, chaos appears as a slow modulation on the oscillation of convective rolls. The time scales, as well as the amplitudes, are well separated, and chaos in each convective roll cannot destroy the spatial coherence.

Coexistence of long-range order with chaos has been discussed recently [36]. A "fero-type" order in a CML with local chaos is found in an Ising-like model [37], while an "antiferro"-like order (or that with a longer wavelength) with local chaos was found in the earlier studies of CML [2][4]. The long-ranged order in the present model belongs to the latter example, where long-range spatial order is maintained by the separation of scales between the collective motion and chaos. As for the shape of Lyapunov spectra, there is one difference between the earlier studies and the present one. In the former, a stepwise structure is often seen [4], while the spectra are rather smooth (with an almost linear slope) in the present case (see figure 17). We believe that this distinction is due to the fact that the domain structure is rigid in the former [4], while the boundary of each roll is not so rigid.

The amplitude of the oscillation of the rolls increases with $\Delta T$, until oscillatory bursts with a large amplitude appear through the interaction of streams of two neighboring cells. Then the collective oscillation loses its stability. Laminar and turbulent states coexist

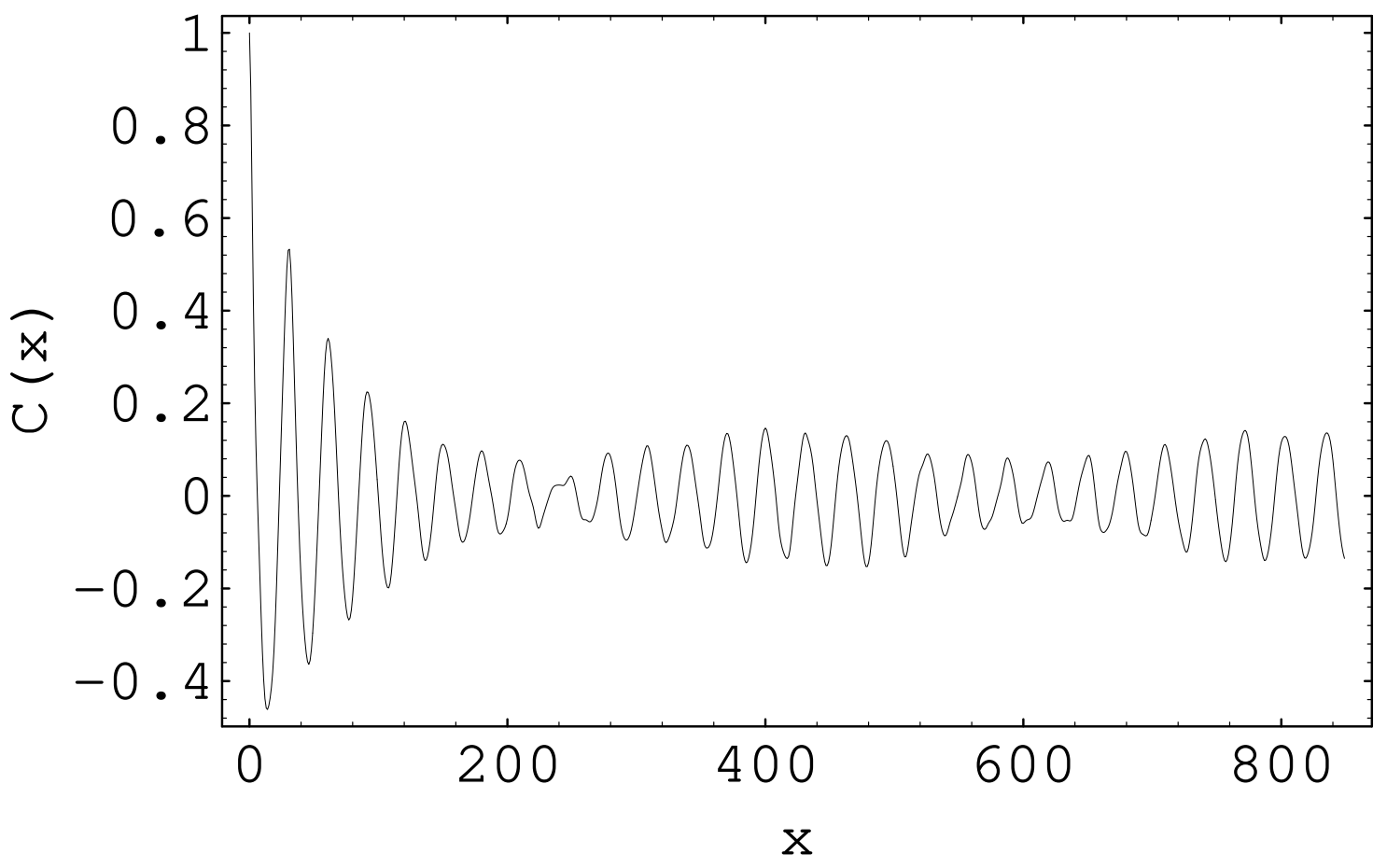

Figure 16: Spatial correlations for coherent chaos

Spatial correlation function at coherent chaos, by the average of $10^4$ time steps after discarding $10^4$ transients. There are many positive Lyapunov exponents here, indicating high dimensional chaos, while the spatial structure still remains. $\lambda = 0.02, \nu = \eta = 0.2, N_x = 17, \Delta T = 0.05$, $\Gamma = 100$.

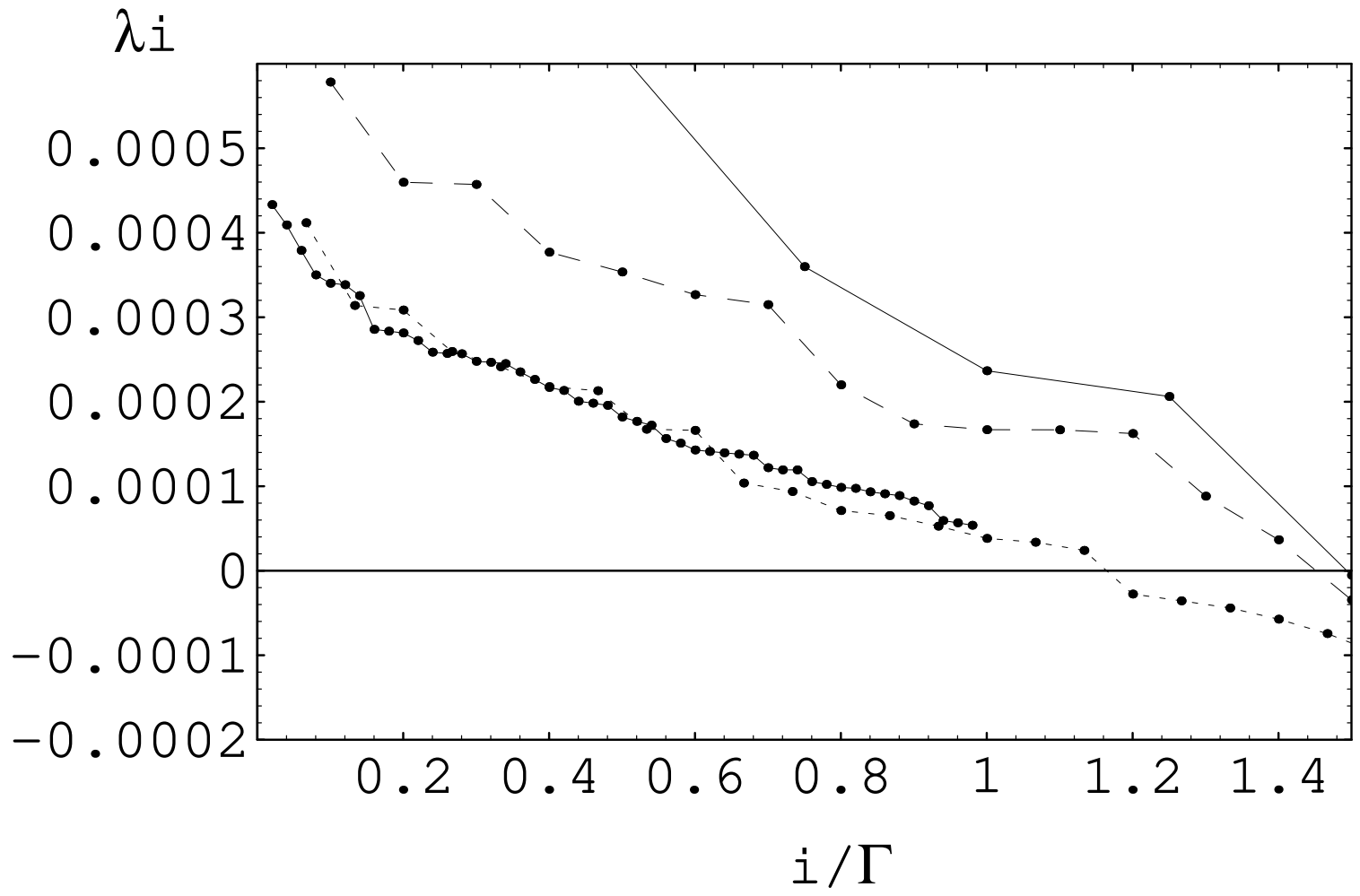

Figure 17: Scaled Lyapunov spectrum

At an intermediate value $\Delta T$ for the collective oscillation (below the STI), chaos with spatial coherence has been observed. The figure shows ordered Lyapunov exponents $\lambda_i$ versus $i/\Gamma$. $\lambda = 0.02, \nu = \eta = 0.2, N_x = 17, \Delta T = 0.05$, $\Gamma = 4$ (thin solid line), 10 (broken line), 15 (dotted line), 50 (solid line). In the computation, we have obtained only the first 50 exponents. The spectra for $\Gamma = 15$ and 50 agree rather well by the scaling of $i/\Gamma$.

in the space time diagram, which leads to spatiotemporal intermittency. The collective oscillatory motion here is a prelude to spatiotemporal intermittency.

During this cooperative oscillation, we have also observed travelling waves. The rolls move to the left or right (depending on the initial conditions) with the oscillation. These travelling waves have been observed in experiments [38], as well as in a simple CML model [39]. In our simulations, we have found that several attractors coexist which correspond to different traveling speeds. The details of the travelling wave will be reported elsewhere.

# 7 Spatiotemporal Intermittency

In this section, we study spatiotemporal intermittency as a standard route to spatiotemporal chaos. The transition to spatiotemporal chaos is rather different from that to temporal chaos. In spatially extended systems, the most well known transition to spatiotemporal chaos occurs through spatiotemporal intermittency (STI). STI occurs through the propagation and connection of chaotic bursts within the laminar domains. There the system consists of a mixture of laminar domains and chaotic bursts. STI was first studied in (diffusively) coupled map lattices for some classes of local maps [3], and has also been observed in partial differential equations [36]. Critical properties were studied in detail in connection with directed percolation [36][40].

In Rayleigh-Bénard convection, STI has been observed in systems with large aspect ratios [40][41][42], and has been investigated intensively. Other examples of STI are observed in electric convection of liquid crystals [43][44], rotating viscous fluids [45], and so on [46]. STI is now believed to form a universality class for the transition to turbulence in spatially extended systems.

In this section, we study this STI transition to spatiotemporal chaos in a system with a large aspect ratio, mostly by fixing it to $\Gamma = N_x/N_y = 50$ and adopting a periodic boundary condition for the horizontal direction. We also discuss the Prandtl number dependence of the transition.

Below the onset of STI ($\Delta T_{STI}$), all the rolls have the same frequency and wavelength. With the increase of $\Delta T$, the spatial coherence of the rolls' oscillations is gradually lost. To see this change, we have measured the distribution of the local wavelength, which is estimated as the distance between two local maxima of $v_y(x, N_y/2)$. For $\Delta T < \Delta T_{STI}$, this distribution has a sharp peak at a single value, while the peak gets broader and broader as $\Delta T$ increases beyond $\Delta T_{STI}$ (figure 18).

A typical example for the space-time positions of the convective rolls in STI is shown in figure 15(b). Two regions coexist; the "turbulent" ones where the spatial periodicity is lost, and the "laminar" ones with local coherence of the rolls' oscillations.

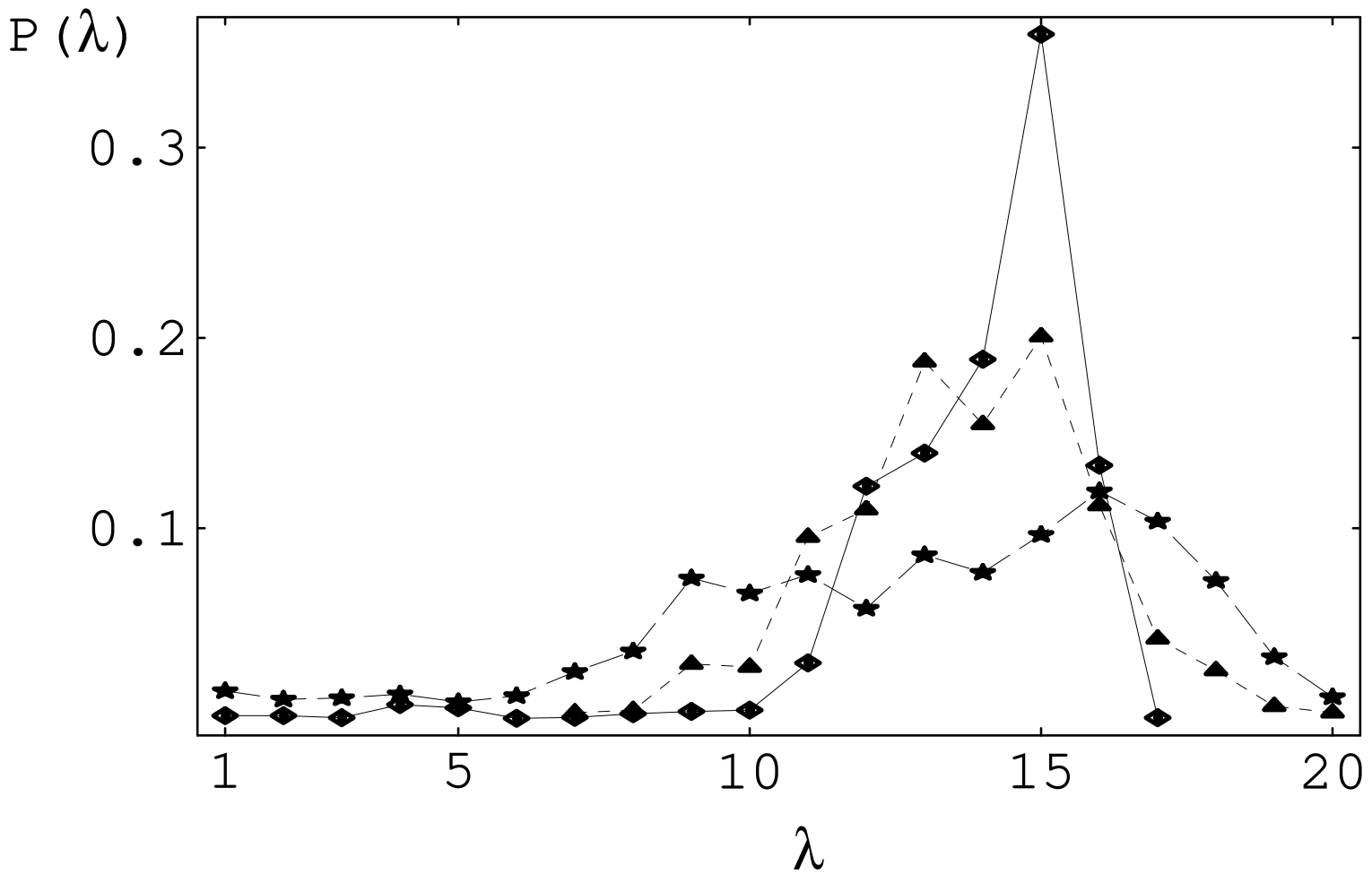

Figure 18: The probability distribution of the local wavelength

For low $\Delta T$, the rolls oscillate coherently (there are no turbulent patches in space and time), and the distribution of local wavelengths has a sharp peak. The peaks gets broader with the increase of $\Delta T$ due to the turbulent patches. Each distribution is taken over $10^7$ time steps sampled per 100 time steps after 1000 initial transients. $\lambda = 0.02, \nu = \eta = 0.2, N_x = 17, \Gamma = 50, \Delta T = 0.01$ (solid line), 0.02 (dotted line), 0.03 (broken line).

To distinguish the laminar and turbulent regions numerically, we have adopted the following criterion with the use of the local wavelength of the convective rolls; By introducing the mean wavelength $\lambda_0$ at the onset of STI and a given tolerance zone $\Delta\lambda$, we assume that the behavior at a position is laminar if the local wavelength there (i.e. the size of the cell) satisfies $\lambda_0 - \Delta\lambda < \lambda$; otherwise it is called turbulent [4]. By using this binary representation, the spatiotemporal diagram for STI is plotted in figure 19. The fraction of the turbulent patches increases with $\Delta T$, which are connected in spacetime near the onset of STI.

STI has been studied both experimentally and numerically. Following previous studies [3][36], we quantitatively characterize the STI behavior, with the use of the distribution $P(L)$ of the laminar domains of length $L$. The existence of two different regimes can clearly be observed. At the onset of STI, the distribution shows a power law (figure 20).

[4]The following results do not depend on the choice of $\Delta\lambda$ provided $3 < \Delta\lambda < 9$ is satisfied.

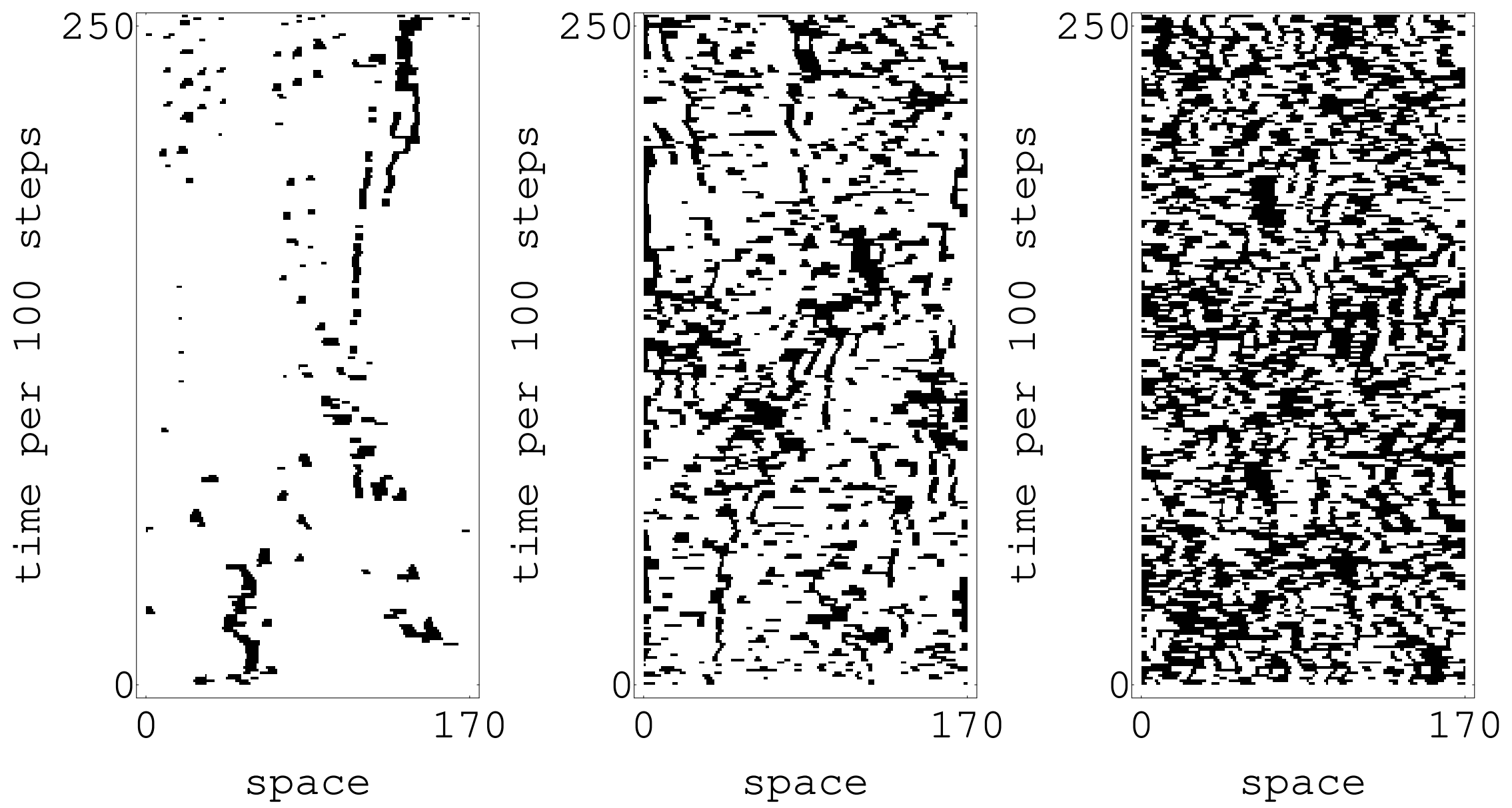

Figure 19: Binary representation for STI

Space time diagram for STI. A black pixel means a turbulent region defined by the criterion in the text. The fraction of turbulent patches increases with $\Delta T$. $\Gamma = 10$, Left: $\Delta t = 0.04$, Center: $\Delta T = 0.12$, Right: $\Delta T = 0.30$. The other parameters are same as in figure 18.

The exponent of this power is $2 \pm 0.2$, and agrees with that found in experiments. Beyond the onset of STI, the distribution is exponential (figure 21), and is fitted by

$$P(L) = C \exp(-\frac{L}{L_{STI}}). \tag{12}$$

It is found that the inverse of characteristic length $1/L_{STI}$ is proportional to the temperature difference $\Delta T$ (figure 22). Furthermore this $\Delta T$ dependence of $L_{STI}$ is invariant against a change of $\lambda$ from 0.02 to 0.1.

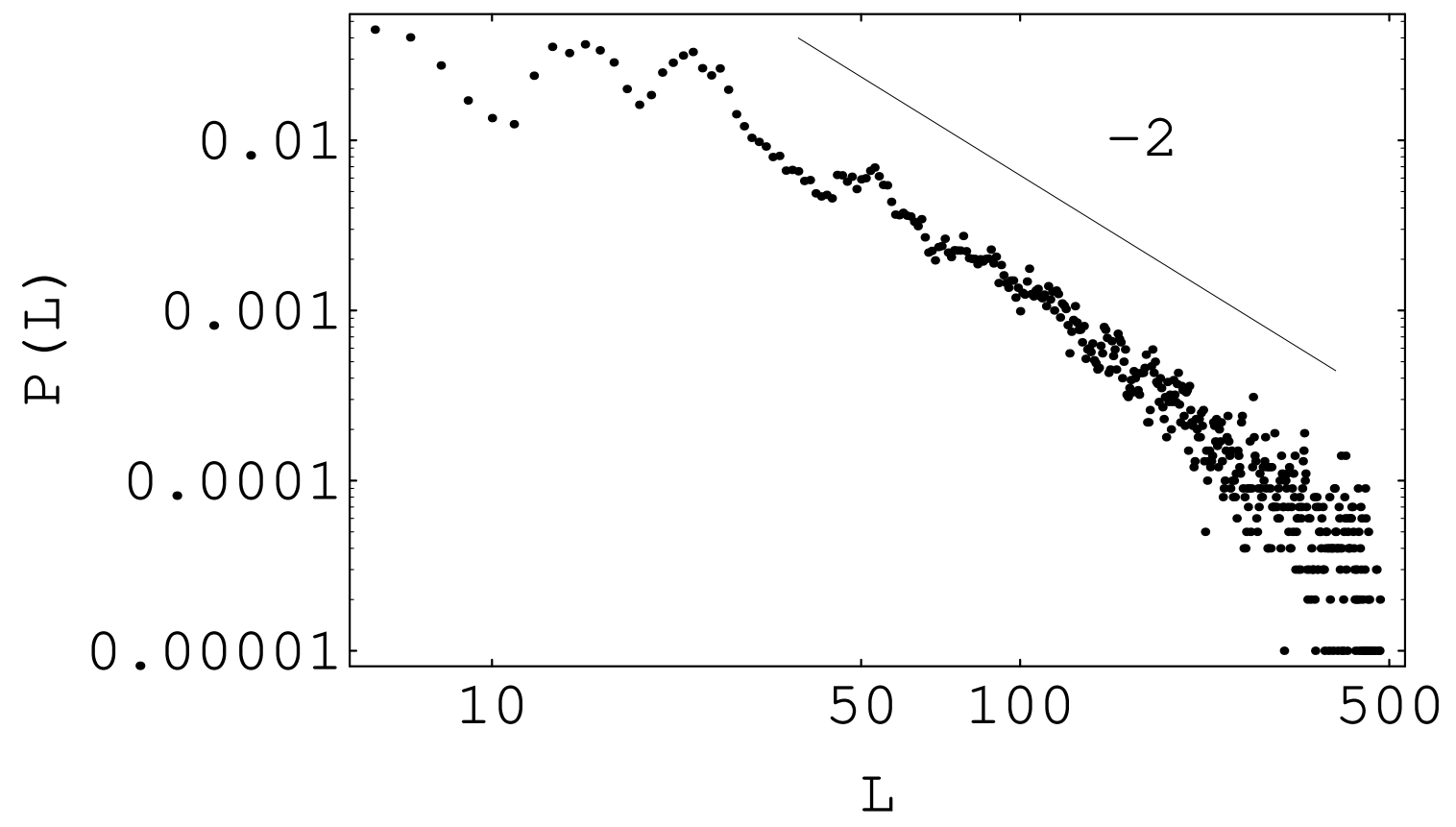

Figure 20: Log-log plot of the distribution of the lengths of the laminar domains

The distribution of the lengths of the laminar domains is plotted by sampling $10^5$ steps, starting from a random initial condition with $\Delta T = 0.05, \lambda = 0.02, \nu = \eta = 0.2, N_x = 850, N_y = 17$. Near the onset of STI, the distribution of the length of laminar domains obeys the power law, with the exponent $2 \pm 0.2$.

We have also measured the spatial correlation function for the vertical velocity (see equation (11)) This correlation function $C(x)$ oscillates with $x$ because of the existence of the roll structure, whose amplitude decays with $x$. The absolute value of the local maxima and minima for $C(x)$ shows exponential decay. We fit $C(x)$ above the onset of STI as

$$C(x) = C_0 \exp(-x/\xi_{STI}). \tag{13}$$

The inverse of the correlation length $1/\xi_{STI}$ increases with $\Delta T$ as is shown in figure 24. This divergence of the correlation length is common in the STI transition, although it is not easy to estimate the value of the exponent accurately. The divergence is a consequence of the laminar domains in figure 22.

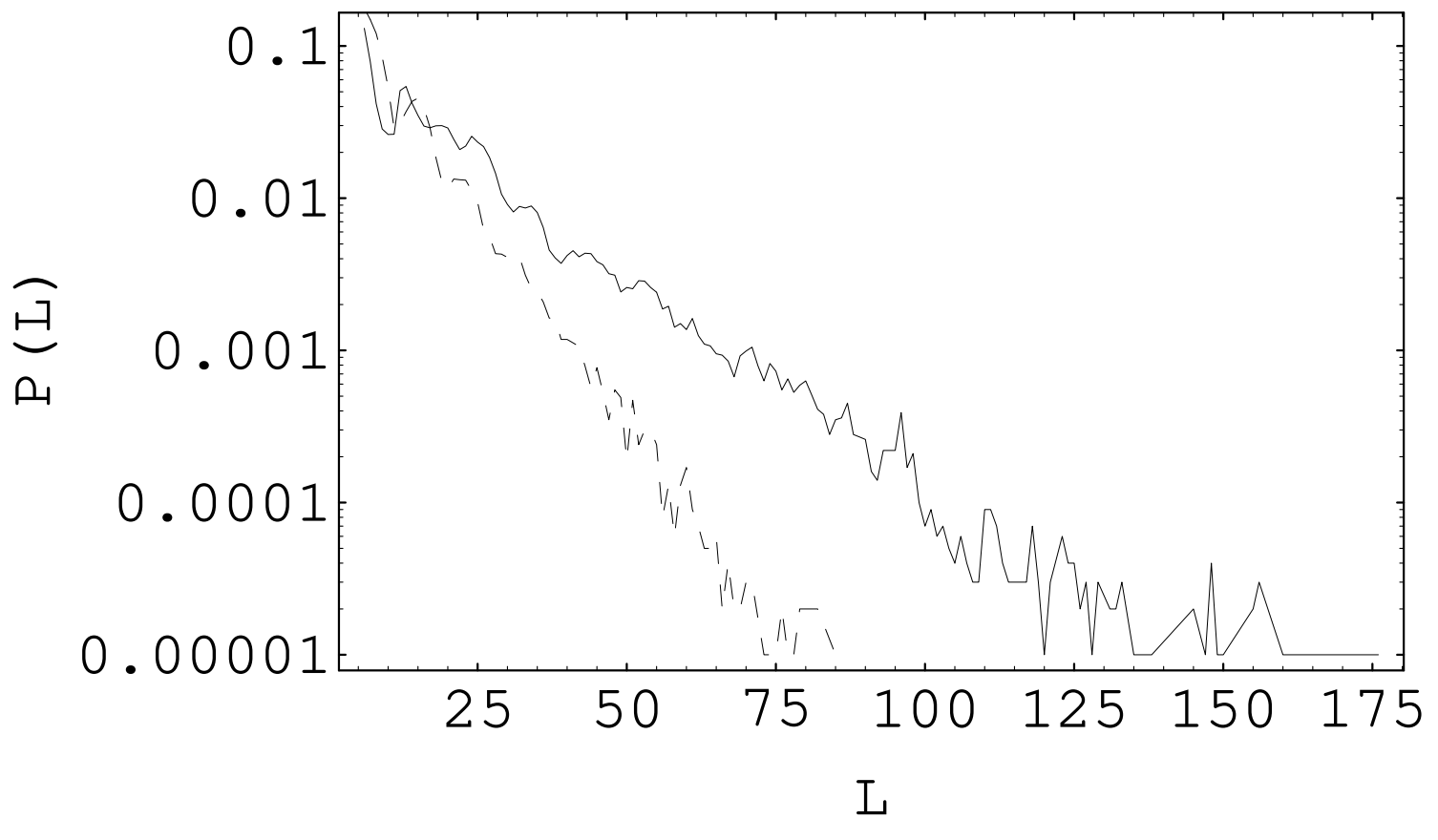

Figure 21: Semi-log plot of the distribution of the lengths of the laminar domains

Semi-log plot of the distribution of laminar regions, for $\Delta T = 0.1$ (solid line) , and 0.5 (broken line). The other parameters are same as in figure 20. Above the onset of STI, the distribution obeys the exponential form ($\Delta T_{STI} \sim 0.05$), whose decay rate decreases to zero as $\Delta T$ approaches $\Delta T_{STI}$.

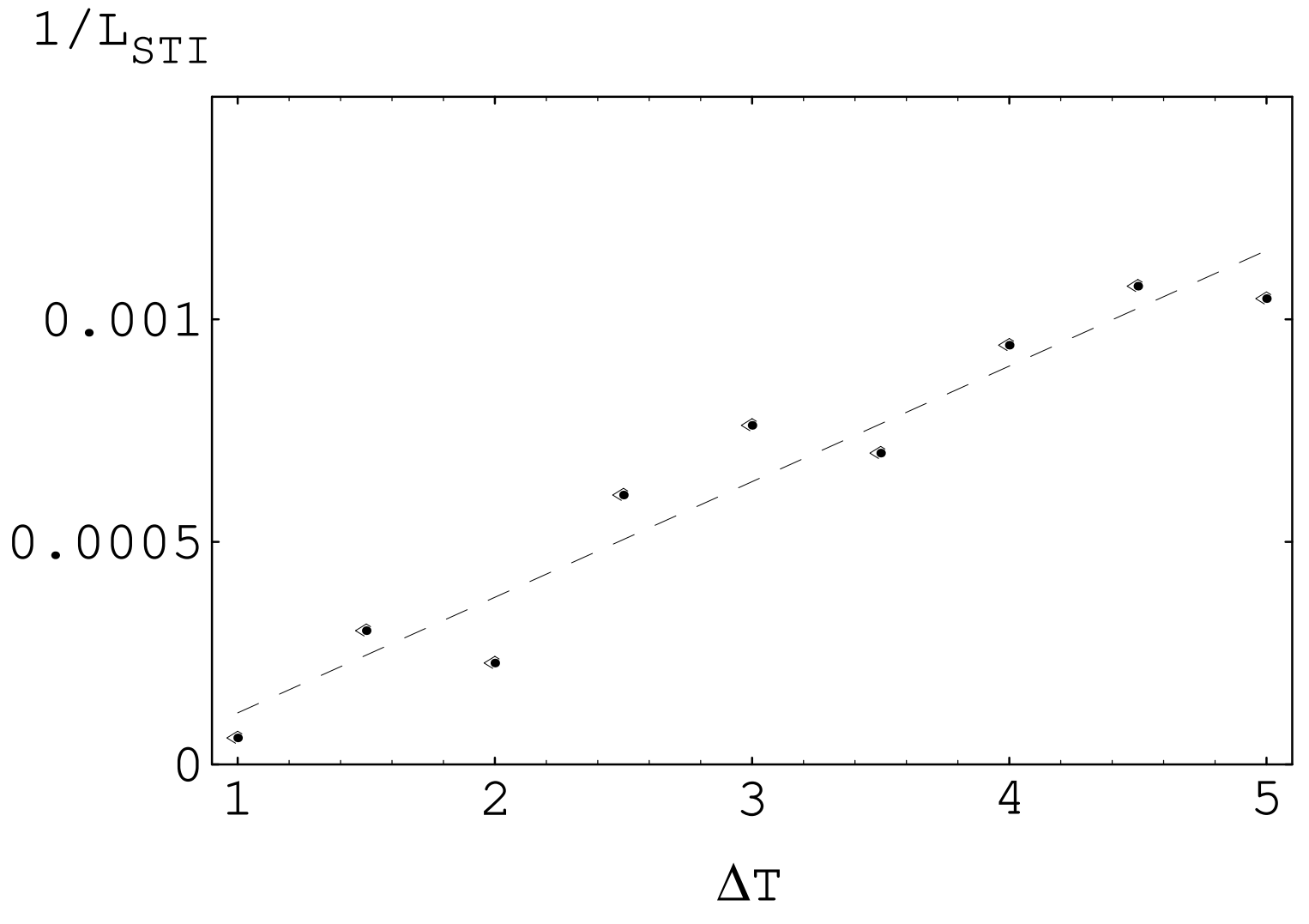

Figure 22: $\Delta T$ dependence of the spatial characteristic length

The inverse of the characteristic length $1/L_{STI}$ of the distribution is plotted versus $\Delta T$. Each dot is obtained by fitting the distribution of the laminar domains with equation (12). The parameters are same as in figure 20.

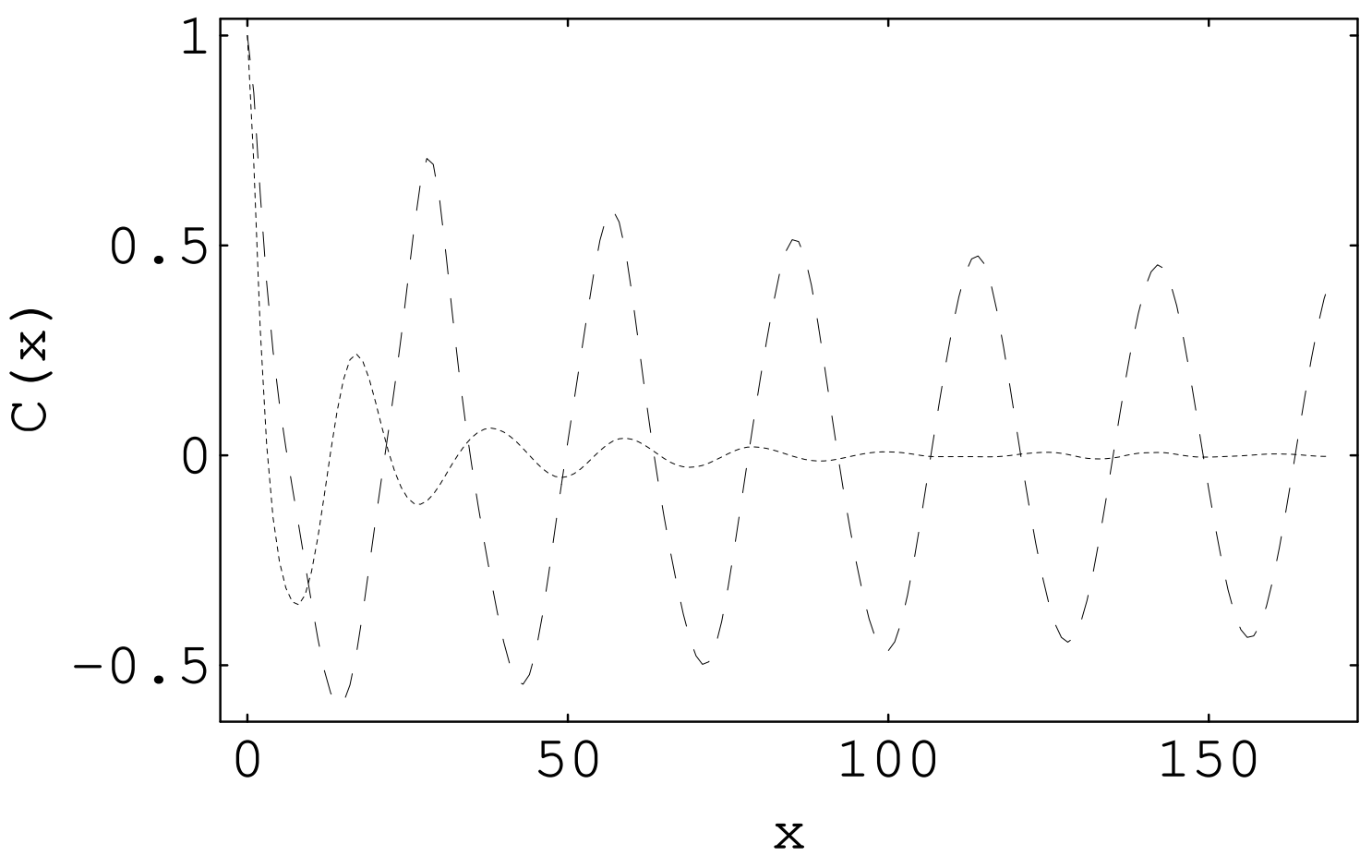

Figure 23: Spatial correlation function

The spatial correlation function $C(x)$ is plotted by the average over $2.0 \times 10^6$ time steps (see text for the definition of $C(x)$), for $\Delta T = 0.1$ (broken line) 0.5 (dotted line). The characteristic length of the decay of the correlation decreases with the increase of $\Delta T$. The other parameters are same as in figure 20; $N_y = 17, N_x = 340$.

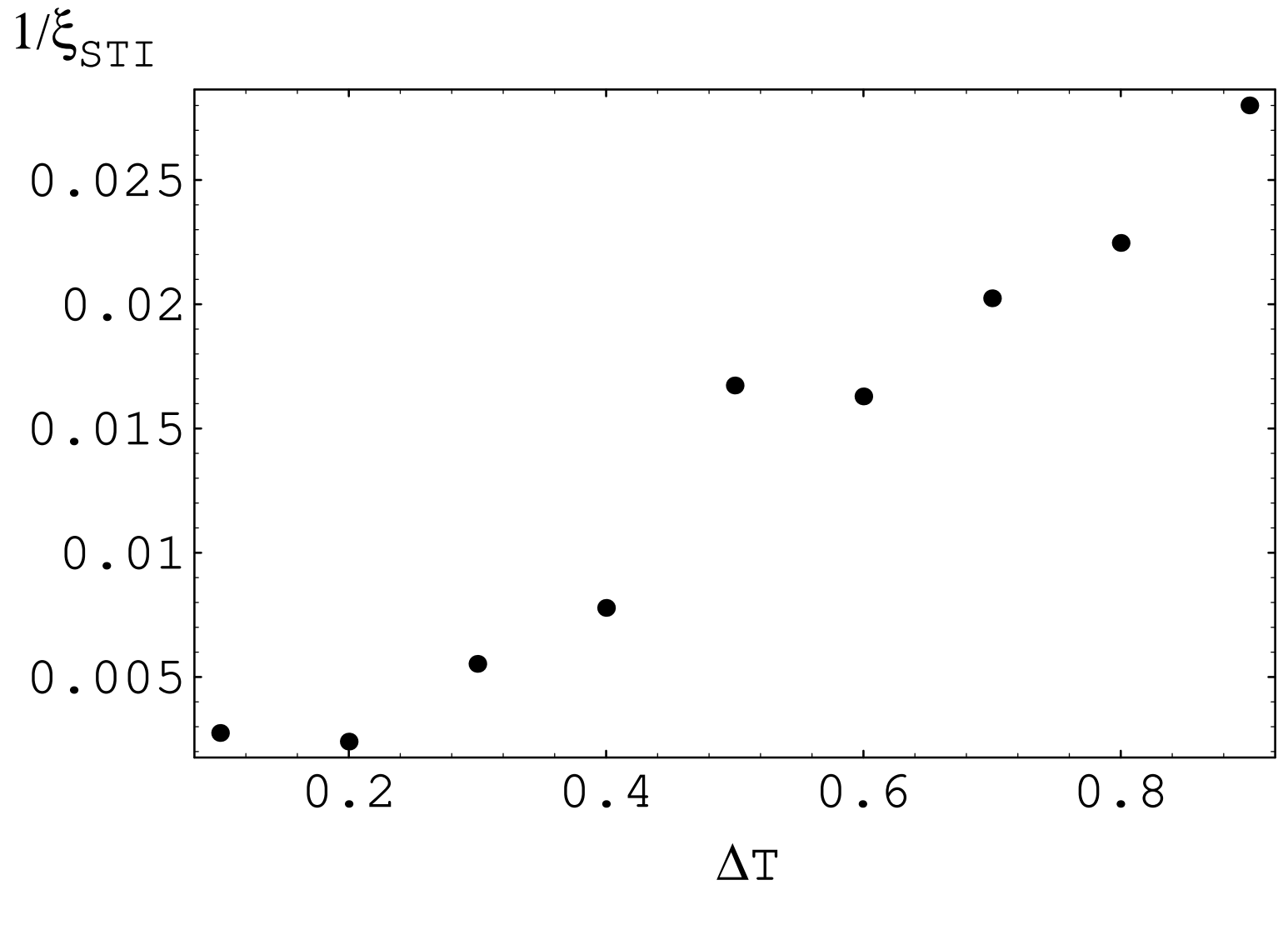

Figure 24: Rayleigh number dependence of the spatial correlation length

The best fit value of $\xi_{STI}$ in equation (13) is plotted versus $\Delta T$. The correlation length $1/\xi_{STI}$ increases with $\Delta T$.

In order to find the Prandtl number dependence near the onset of STI, we define the following characteristics

$$F = \langle \frac{\text{number of turbulent patches}}{\text{number of patches}} \rangle. \tag{14}$$

where $\langle \cdots \rangle$ denotes the temporal average. A global characterization of STI is given by the evolution of the turbulent fraction $F$; $F$ is calculated as the averaged total length occupied by the turbulent cells, divided by the length of the container. By plotting this turbulent fraction versus the thermal conductivity $\lambda$, we find that the critical value $\Delta T_{STI}$ almost linearly increases with the Prandtl number. The power law behavior and its exponent at the onset of STI are invariant under a change of of Prandtl number.

We have also calculated the Lyapunov spectra by changing $\Delta T$ and the aspect ratio. In and above the STI region, the ordered Lyapunov exponent decreases almost linearly with its index (see figure 25). Neither a plateau at the null exponent nor a stepwise structure is observed. This linear shape is distinguished from that of type-I STI in some coupled map lattices [3], and is consistent with that for the type-II STI of the coupled logistic lattice [4]. With the increase of the aspect ratio, the scaled Lyapunov exponents $\lambda_i$ versus $i/\Gamma$ approaches a single form, as is shown in figure 25. The approach is rather slow near the onset of STI, due to the long range spatial correlation, where the intermittent appearance of large laminar patches enhances the statistical fluctuation of the Lyapunov exponents.

# 8 Transition From Soft to Hard Turbulence

When $\Delta T$ is increased further, the roll patterns collapse, and the convection shows turbulent behavior. In recent experiments, Libchaber's group has found a transition in turbulence. The phase at lower $\Delta T$ is called soft turbulence, while the latter at higher $\Delta T$ is called hard turbulence [47][48]. They have characterized this transition by the temperature distribution in the middle of container. According to their experiments, the distribution is Gaussian in the soft turbulence regime, while it is exponential in the hard turbulence regime. They have also pointed out that the transition is due to the destruction of the boundary layer and the formation of hot and cold plumes.

Let us discuss this soft/hard turbulence transition in our model. By increasing $\Delta T$, hot and cold plumes start to appear (figure 26), above some transition temperature. Plumes in our model are defined as isolated sets of few connected lattice points with larger or smaller energy $E$ than their neighbors. Slightly above the transition temperature for the plume formation, a hot plume cannot reach the top plate (and vice visa for a cold plume). The

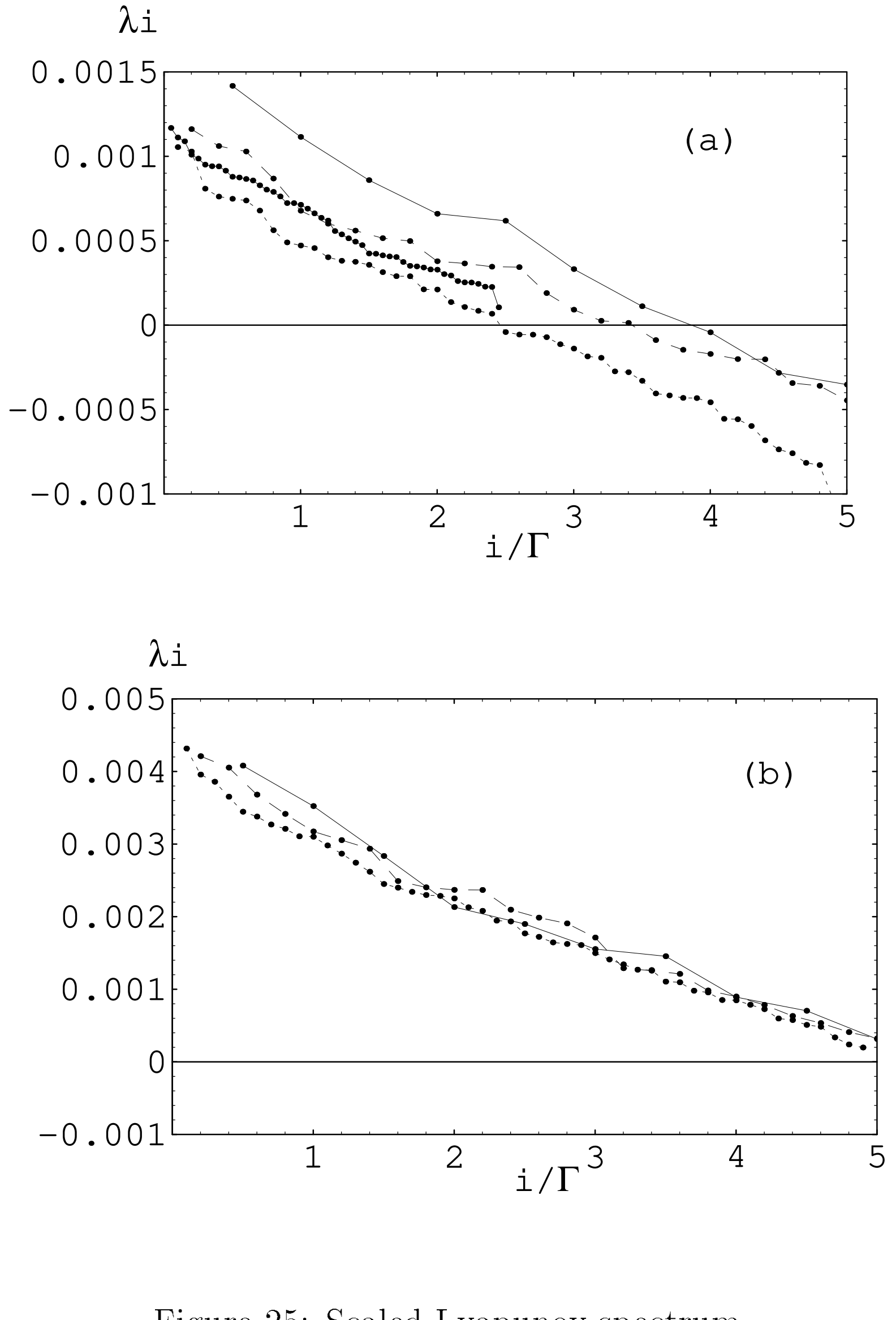

Figure 25: Scaled Lyapunov spectrum

The ordered Lyapunov exponents $\lambda_i$ versus $i/\Gamma$ are plotted. $\lambda = 0.02, \nu = \eta = 0.2, N_x = 17$. (a) $\Delta T = 0.1$ (near onset of STI), $\Gamma = 2$ (thin solid line), 5 (broken line), 10 (dotted line), 20 (solid line). (b) $\Delta T = 0.5$, $\Gamma = 2$ (thin solid line), 5 (broken line), 10 (dotted line). In the computation, we have obtained only the first 50 exponents by the average of $10^4$ time steps after discarding $10^4$ transients. The Lyapunov spectrum approaches a single form with the increase of aspect ratio. Above the onset of STI (b), this convergence is faster than for (a).

boundary layers are still preserved. With the further increase of $\Delta T$, plumes can reach the opposite plate, breaking the boundary layers. This observation agrees with the picture by Libchaber's group for the transition between the soft (for former) and hard turbulence [47][48].

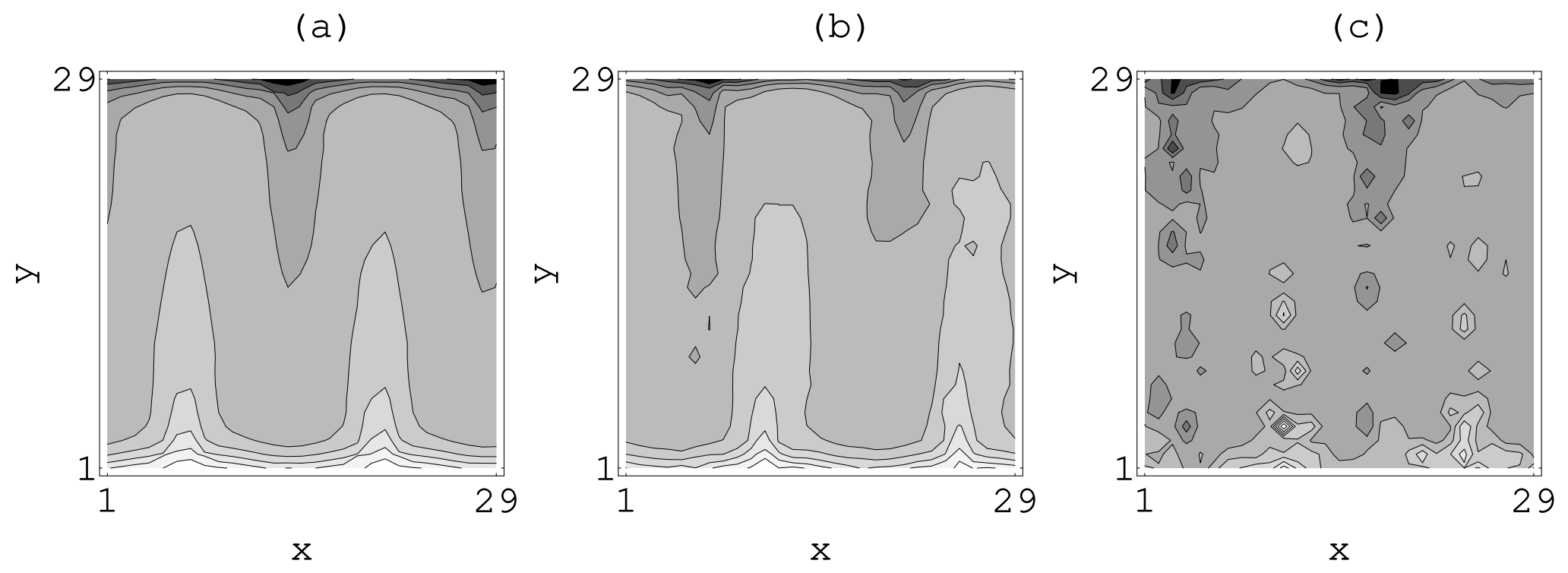

Figure 26: Contour plot of temperature field

The snapshots of equi-temperature lines are plotted for (a)$\Delta T = 1.0$ (b)$\Delta T = 3.0$, and (c)$\Delta T = 10.0$. In (b) (in the soft turbulence region), plumes exist near the boundary layer. In (c) (in the hard turbulence region), plumes can reach the opposite boundary, and the boundary layer is destroyed by these plumes. $\lambda = 0.4, \nu = \eta = 0.2, N_x = 29, N_y = 29$

To confirm the transition quantitatively, we have measured the distribution of $E(x, N_y/2)$, by sampling over a given time interval. As is plotted in figure 27, the distribution shows the transition from Gaussian to exponential, in agreement with experiments.

To characterize the change of the distribution quantitatively, we have also calculated the flatness

$$f = \langle (E - \langle E \rangle)^4 \rangle / \langle (E - \langle E \rangle)^2 \rangle^2. \tag{15}$$

At low Prandtl numbers, the flatness rises from 3 to 6 with the increase of $\Delta T$, while it rises continuously to 12 at high Prandtl numbers. Moreover, the plateau around the flatness 3 (in the soft turbulence region) gets narrower by increasing the Prandtl number (figure 28). This Prandtl number dependence of the flatness is our prediction here, which should be confirmed by experiments in the future. Our CML provides the first simple model for the soft-hard turbulence transition[49]. Our observation of the energy pattern (figure 26) also suggests that this transition is associated with the percolation of plumes at the bottom plate.

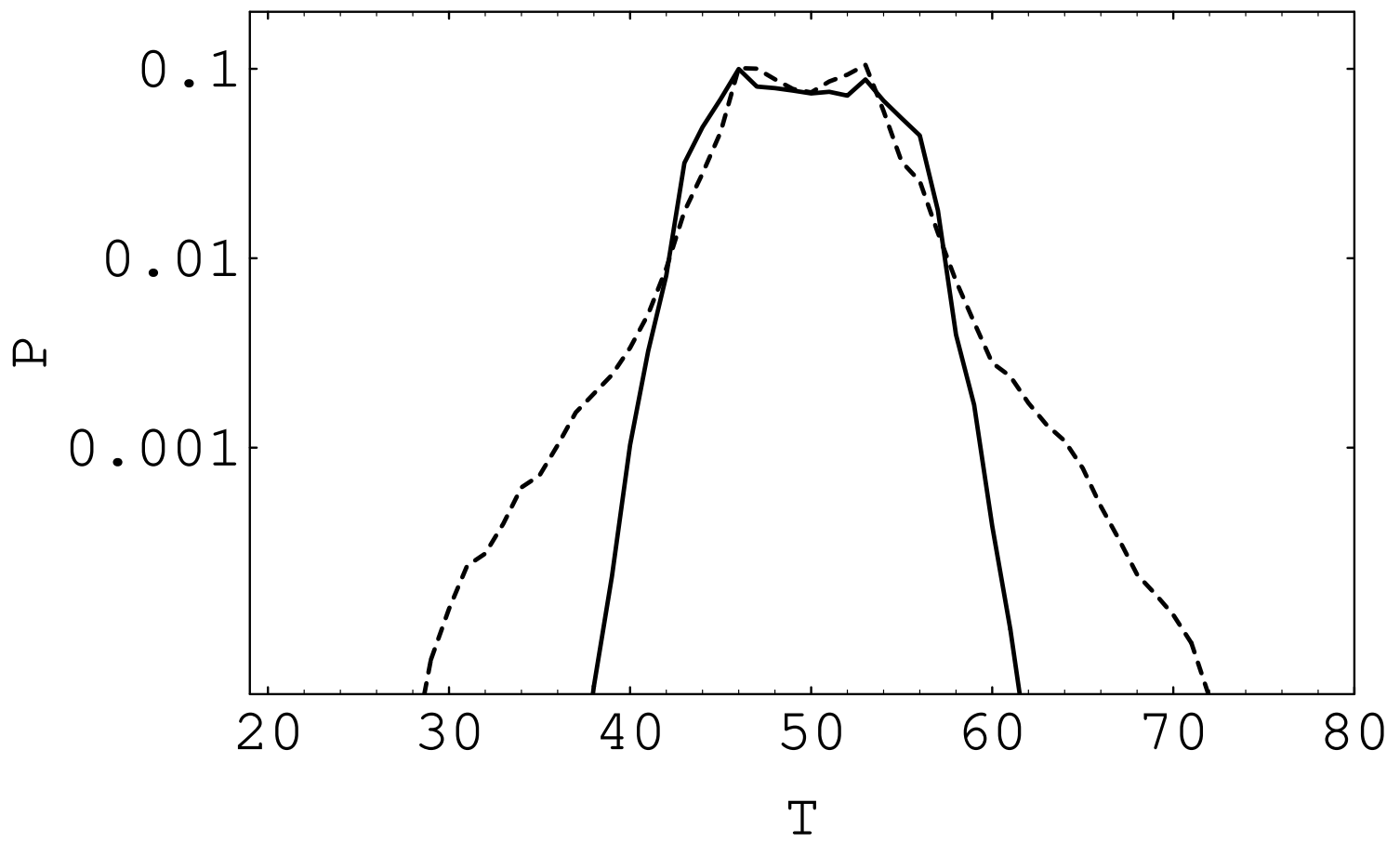

Figure 27: Distribution function of the temperature

The distribution of the temperature $E(x, N_y/2)$ in the middle of the container, measured from the histogram of the temperature sampled over $10^5$ time steps. The distribution changes its form from Gaussian to exponential, indicating the soft and hard turbulence respectively. $\lambda = 0.4, \nu = \eta = 0.2, N_x = N_y = 29$, solid line:$\Delta T = 3.0$, dotted line:$\Delta T = 5.0$.

We have also varied $\eta$, which expresses the pressure effect, from 0.2 to 0.4. The flatness for the temperature distribution scatters around from 2.6 to 3.0 in the soft turbulence region ($\Delta T = 3.0$), without any systematic deviation from the Gaussian shape. Thus our transition is a robust property against the change of $\eta$, which is important for the justification of our approach, since $\eta$ represents a rather artificial term in our modeling.

To study the transition in terms of dynamical systems, the Lyapunov spectrum and Kolmogorov-Sinai entropy are computed (figure 29). By increasing $\Delta T$, the number of positive Lyapunov exponents also increases, in contrast with the chaotic itinerancy case discussed in § 5 (figure 13). Within our simulation, no plateau at the null exponent is clearly visible. At present it is not sure if this lack of the plateau implies the absence of the cascade process, or it is just because the number of lattice points is not large enough.

# 9 Pattern Formation

Extension of our model to three dimensions is quite straightforward. We have simulated three-dimensional convection in rectangular and cylinderical containers, taking a fixed boundary at the wall. Here the pattern formation of convective rolls requires a long time, due to slow motion of defects between locally aligned rolls. Temporal evolution of roll

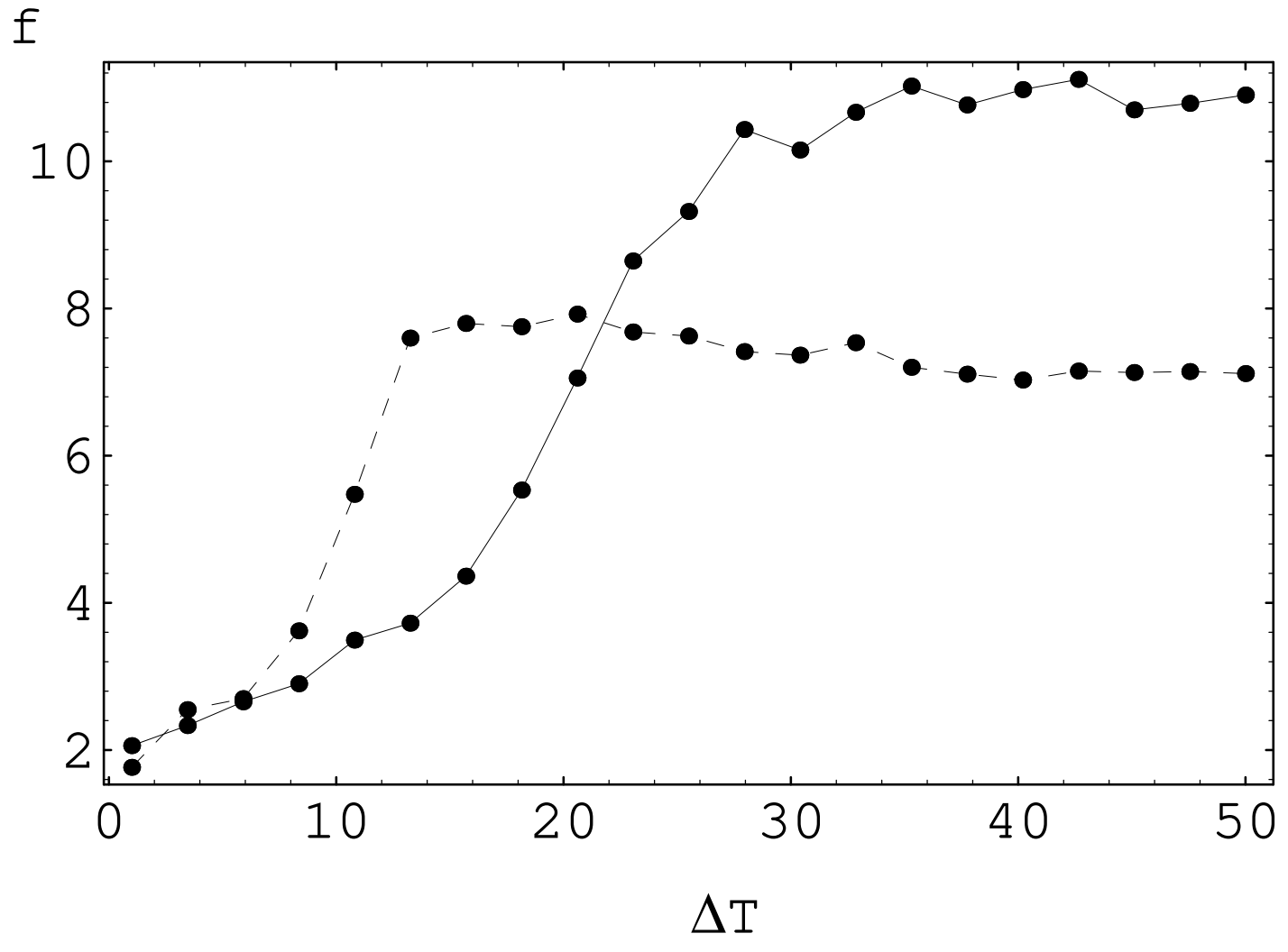

Figure 28: The flatness of the temperature distribution

At low Prandtl numbers, the flatness of the distribution increases with $\Delta T$ and saturates around 6.0, while at high Prandtl numbers, it increases until 12. line: $\lambda = 0.404$, dotted line: $\lambda = 0.116$. Each dot is obtained from the average over 20000 time steps. The other parameters are same as in figure 27.

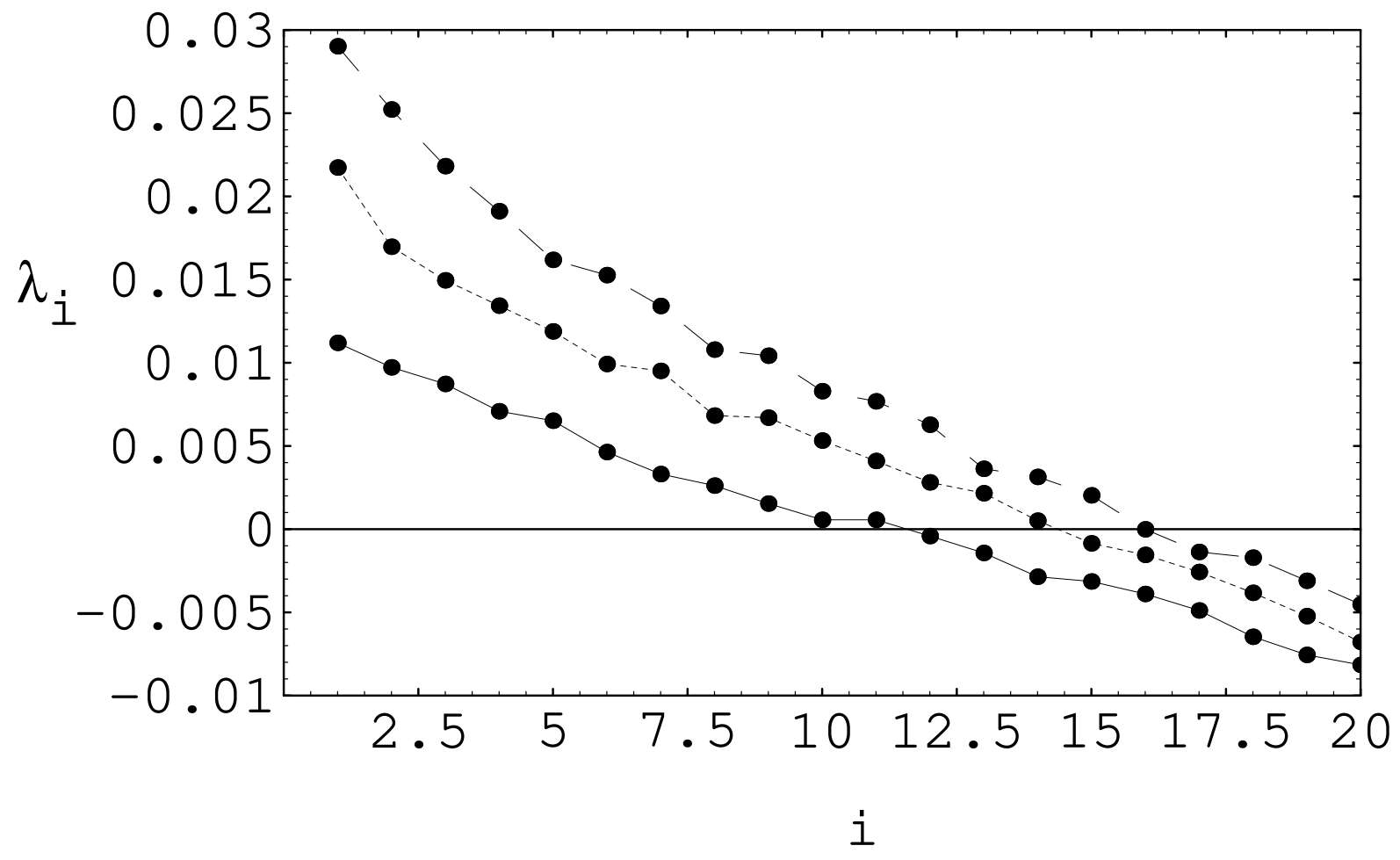

Figure 29: Lyapunov spectrum for the turbulent regime

The first 20 Lyapunov exponents, computed by the average over $10^4$ time steps. In the turbulent regime, the number of positive Lyapunov exponents increases rapidly, which should be compared with the chaotic itinerancy motion in § 5. $\lambda = 0.4, \nu = \eta = 0.2, N_x = N_y = 30$, solid line: $\Delta T = 2.0$, dotted: 3.0, broken: 5.0.

patterns is given in figure 30, which shows clear agreement with experimental observations [50]. Starting from an almost homogeneous field, rolls are formed locally within a short time, while defects between rolls move slowly. The domain size of aligned rolls increases so slowly that the irregular motion of the defects remains over many time steps. If $\Delta T$ is larger, these defects form cellular structures as in figure 30(f).

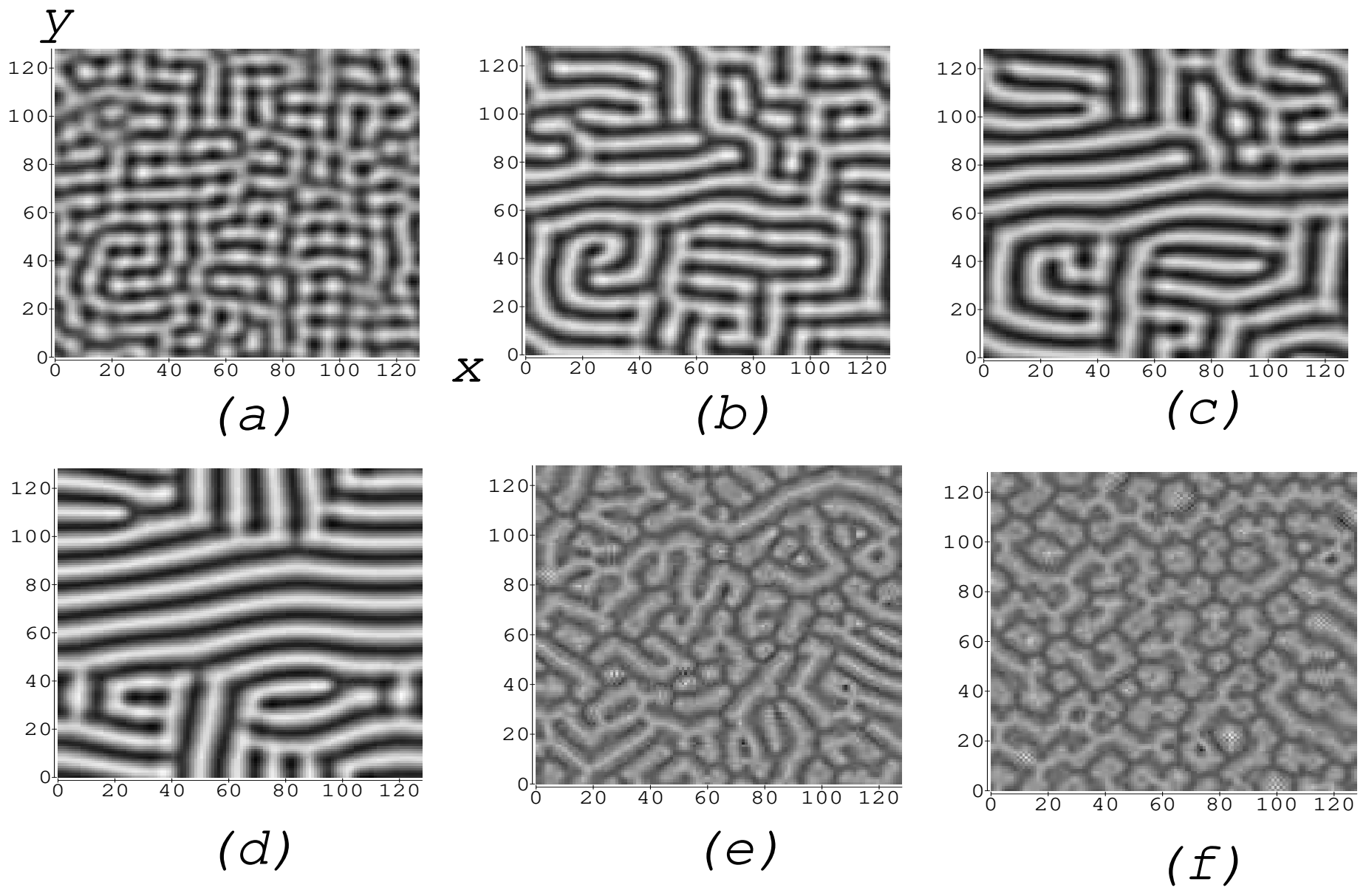

Figure 30: Pattern formation of convective rolls

Roll pattern for three-dimensional convection. Snapshot of the perpendicular velocity $v_z$ at the middle plate $(x, y, N_z/2)$ is shown with the use of gray scales. The lattice size is $(N_x, N_y) = 125 \times 125$ (horizontal), and $N_z = 9$. $\nu = \lambda = 0.2$. Random initial condition were used. $\Delta T = 0.6$ and (a) time step 500 (b) 1000 (c) 2000 (d) 5000. $\Delta T = 2.0$ and (e) time step 50 (f) 5000.

To see the pattern formation process, we have measured the spatial power spectrum $P(k)$ of the vertical velocity $v_y(x, N_y/2)$ for the 2-dimensional model;

$$P(t,k) = \langle | \int_0^{N_x} dx \; v_y^t(x, N_y/2) \exp(-i(kx)) |^2 \rangle \tag{16}$$

where $\langle \cdots \rangle$ means the sample average over different initial conditions close to a homogeneous one (see figure 31). Starting from a random initial configuration, convective rolls are locally formed, which leads to the appearance of a peak in the spatial power spectrum. As the pattern formation proceeds, the peak shifts to a lower wave number while it gradually sharpens. We plot the wave number $k_m$ which gives the maximum of $P(t,k)$, versus time $t$ (see figure 32). $k_m$ converges to a characteristic wave number $k_c$ for the stationary state. The figure also shows power law behavior for the pattern formation process. The exponent for the convergent process is $-1/2$ which agrees with experiments as well as the theory of pattern formation. Although the present result is obtained with the 2-dimensional model, we believe that the scaling exponent is invariant in a 3-dimensional case also.

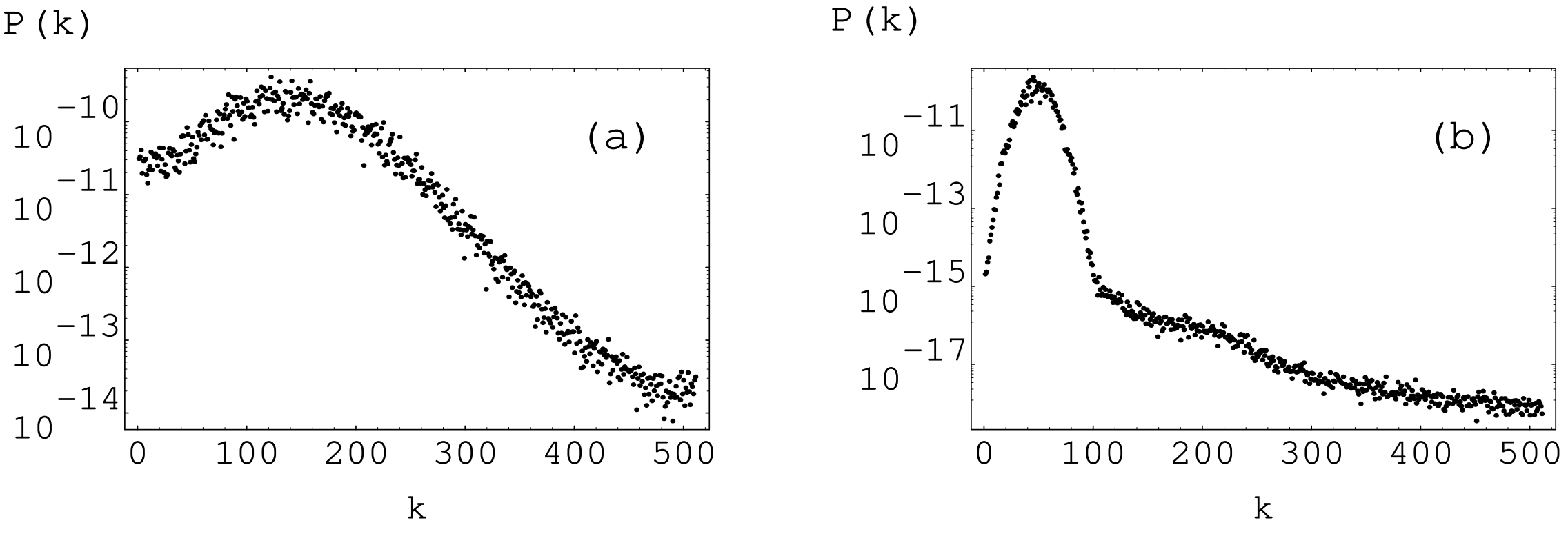

Figure 31: Spatial Power Spectrum

Spatial power spectrum averaged over 10 different initial conditions. (a)$t = 32$, (b)$t = 1024$, $\lambda = 0.2, \nu = \eta = 0.2, \Delta T = 0.01, N_x = 1024, N_y = 17$.

Inclusion of rotation to the convection is rather straightforward. We introduce the centrifugal and Coliolis force procedure before the Lagrange procedure;

$$\vec{v} \longmapsto \vec{v} + 2\vec{\omega} \times \vec{v} + \vec{\omega} \times (\vec{\omega} \times \vec{x}) \tag{17}$$

Here we show only some examples of the spatial patterns (figure 33). By increasing the rotational speed, spiral convective rolls appear. As the rotational speed is further increased, the spiral structure collapses, and a complicated structure is successively formed.

# 10 Summary and Discussions

In the present paper, we have proposed a CML model for Rayleigh-Bénard convection, by introducing a new procedure, i.e., a Lagrangian scheme for the advection. In this procedure, the advective motion is expressed by a quasi-particle.

Our model reproduces almost all phenomenology in convection; formation of rolls and their oscillations, many routes to chaos, spatiotemporal intermittency, transition from soft

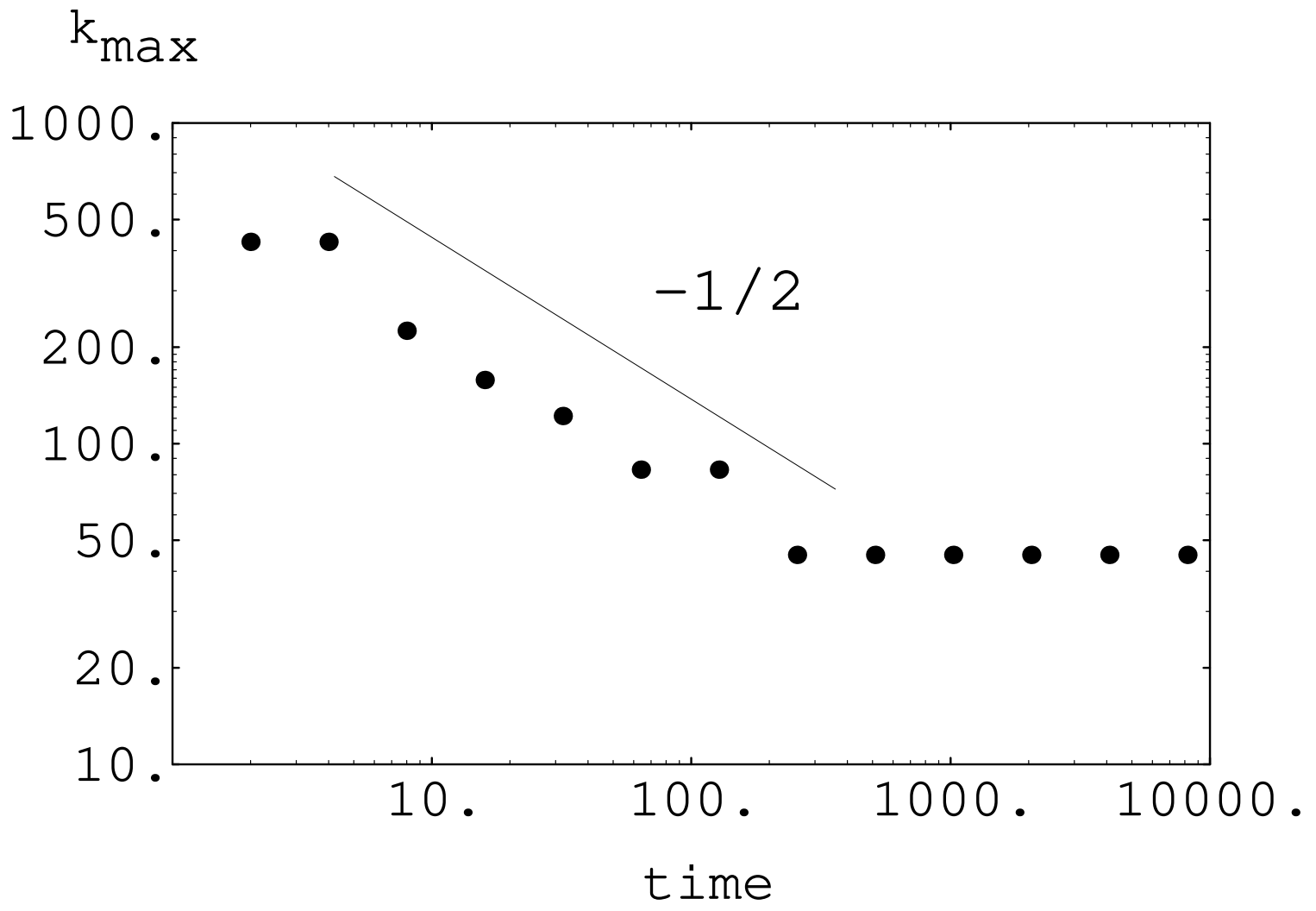

Figure 32: Scaling exponent for the domain growth

Log-log plot for characteristic length (the maximum of spatial power spectrum) versus time. $\lambda = 0.2, \nu = \eta = 0.2, \Delta T = 0.01, N_x = 1024, N_y = 17$

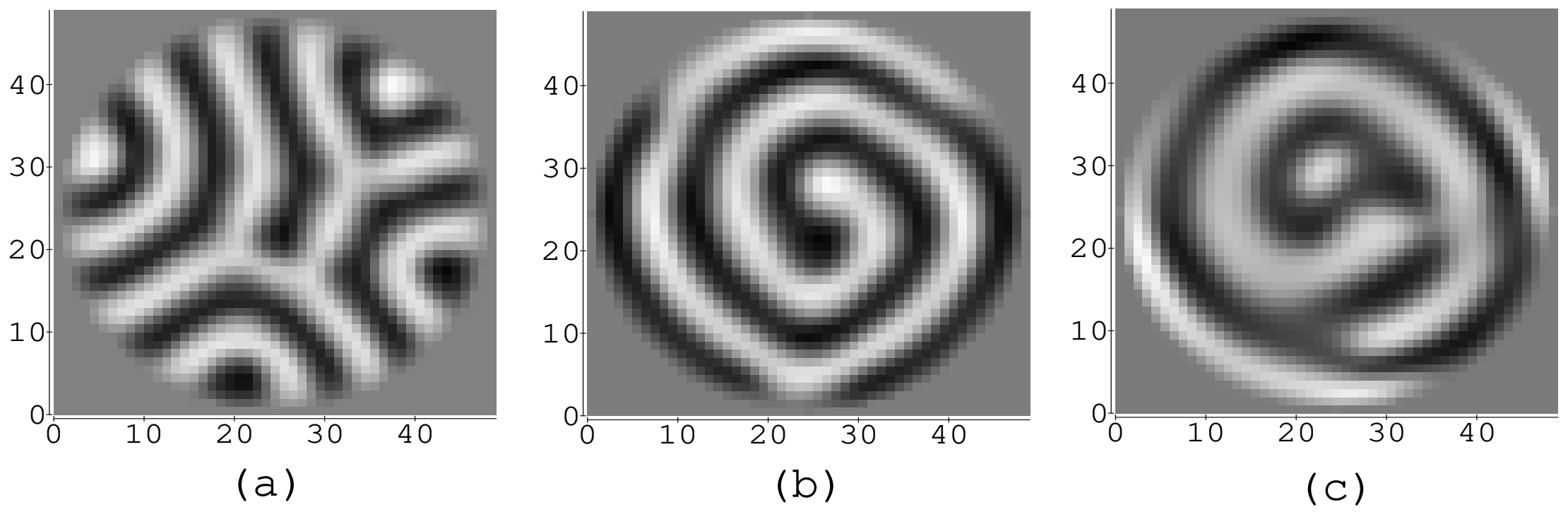

Figure 33: Inclusion of rotation to the convection

Snapshot of the perpendicular velocity $v_z$ at the middle plate $(x, y, N_z/2)$ is shown with the use of gray scales. The lattice size is $(N_x, N_y) = 50 \times 50$ (horizontal), and $N_z = 9$. $\nu = \kappa = 0.2, \Delta T = 1.0$.(a) angular velocity $\omega$ is .001 (b) $\omega = .004$ (c) $\omega = .008$ .

to hard turbulence, and the pattern formation process. Besides qualitative agreement with experimental observations, some quantitative agreements are also obtained; the power law distribution of laminar regions in STI with the exponent $2 \pm 0.2$, and the flatness of the temperature distribution at the soft-hard turbulence transition, in addition to rather trivial agreements on the exponents on the onset of convection, the critical slowing down and the pattern formation. The results on soft-hard turbulence may be the most remarkable, since it provides the first simple model with an agreement on the change of distributions. It is also noted that the role of disconnected plumes is confirmed with the help of the snapshot temperature field.

Furthermore, we have also made several predictions here. (i) In systems with medium aspect ratios, switching between two roll patterns is found which occurs through high-dimensional chaos. At the onset of the chaotic itinerancy, the average lifetime of laminar states diverges. (ii) Spatial long-range order with temporal chaos is found in a system with a large aspect ratio. Spatial correlations do not decay although the number of positive Lyapunov exponents increases with the system size. (iii) For the soft-hard turbulence transition, the calculated flatness of the temperature distribution increases from 3 to 6 at low Plandtl numbers, as is known in experiments. On the other hand it raises till 12 at high Plandtl numbers, which can be checked in future experiments.

Correspondence of our results with experiments is summarized in Table I. Here a dash in the experiment column shows our prediction.

One of the merits of our modelling here lies in the applicability of dynamical systems theory. It is possible to describe the convection phenomena in terms of dynamical systems, in particular by Lyapunov exponents. Collective motion with high-dimensional chaos is thus confirmed, as well as the switch between low- and high- dimensional dynamics at the chaotic itinerancy. Lyapunov spectra for STI and soft/hard turbulence transitions are also obtained.

Some, still, disagree with our CML approach only because our model is not derived from the Navier-Stokes equation. Our standpoint here is that the salient features in convection are irrespective of the details of the models. Such features form universal classes. All of our results suggest that the qualitative features of convection do not depend on the details of the dynamics. This means that our model and real fluid dynamics belong to the same universality class.

One of the advantages of our approach is the possibility to check the robustness of a given feature of convection against the modification and/or removal of processes (see Appendix A). For example, the power law distribution of laminar domains in STI does not depend on the dynamics of the pressure effect, while it crucially depends on the buoyancy procedure. On the other hand, the buoyancy procedure is not relevant to the soft-hard turbulence transition, (but the pressure procedure is). Indeed the distribution change of a passive scalar from a Gaussian to an exponential form is also observed in grid-generated turbulence and stirred fluids. Such universality may be related with the stability against the choice of models.

Table I: Summary of Our Results in Comparison with Experiments, as well as some Predictions

| phenomena | characteristics | CML model | experiment |
|---|---|---|---|
| Onset of Convection | $v_z \sim \epsilon^{\alpha}$ | $\alpha = -1/2$ | $\alpha = -1/2$ |
| Critical Slowing Down | $1/\tau \sim j^{\alpha'}$ | $\alpha' = -1$ | $\alpha' = -1$ |
| Route to Chaos | Qusi-periodic | Hi Plandtl | Hi Plandtl |
| | Period doubling | Low Plandtl | Low Plandtl |
| | Intermittency | Depend on $\Gamma$ | — |
| Chaotic Itinerancy | Lifetime at laminar states | diverges at $\Delta T_{CI}$ | — |
| Coherent Chaos | Spatial Long-range Order with chaotic motion | EXISTS | — |
| Traveling Wave | Coexistence of different speeds attractors | EXISTS | — |
| STI | Distribution of Laminar Domain $P(L) \sim L^{-\gamma}$ (Onset) | $\gamma = 2.0 \pm 0.2$ | $\gamma = 1.8$ |
| Soft/Hard Turbulence | Flatness | 3 to 6 | 3 to 6 |
| | $\langle (E - \langle E \rangle)^4 \rangle / \langle (E - \langle E \rangle)^2 \rangle^2$ | 3 to 12(Hi Prandtl) | — |
| Pattern Formation | characteristic length $\xi = t^{\beta}$ | $\beta = 1/2$ | $\beta = 1/2$ |

Thus our constructive approach is powerful for proposing universal classes of the phenomenology. In our model, for example, the soft-hard turbulence transition is associated with the percolative behavior of plumes. This allocation forms the basis of universality such as the change of the temperature distribution. The essence of the transition does not depend on the details of a model, as long as it belongs to the same universality class.

As we have already discussed in § 2.3, the NS equation is not derived from any microscopic descriptions with rigor. The NS equation is not necessarily the best model for numerical analysis, due to its demand of huge computational resources. In order to globally understand the phenomenology, we must scan over the parameter spaces. Thus fast and interactive computation is important for a model construction. The computational advantage of our model is powerful for predicting universality classes. It should be mentioned that all of our results here have been obtained by workstations. Still, we can reproduce almost all phenomenology in convection, and furthermore we can predict novel classes of behaviors as well as the Prandtl number dependence on the phenomena such as the soft-hard turbulence transition.

Last but not least, it should be mentioned that our Lagrangian procedure is also useful to construct a CML for shear flows or Kármán vortices and their collapse. Another important extension of our CML is the inclusion of phase transition dynamics, as is seen in boiling [7] and cloud dynamics. These examples will be reported elsewhere.

## Acknowledgements

We would like to thank M. Sano, S. Adachi, and J. Suzuki for useful discussions, and Fredrick Willeboordse for critical reading of the manuscript and illuminating comments. This work is partially supported by Grant-in-Aids for Scientific Research from the Ministry of Education, Science, and Culture of Japan, and by a cooperative research program at Institute for Statistical Mathematics.

# Appendix

## A "Structural stability" of our Model

In previous sections, we have shown that our model reproduces almost all phenomenology of convection (with some predictions). It is surprising that such a simple model reproduces almost all phenomenology. In this appendix, we discuss the stability of our model to study the "universality" classes of convection. Here, we use the term "universality" in a rather qualitative sense: if a set of models reproduces the same macroscopic properties such as flow patterns and statistical quantities, these models form a "universality class". For example, spatiotemporal intermittency is believed to form such a universality class, since it is observed in a wide range of models with spatial degrees of freedom. Here, we address the following questions. Are there any other models which reproduce the phenomenology of convection? Is a given characteristic also reproduced by modification or removal of some elementary physical processes? In other words, are macroscopic properties robust against the structural change of models?

The coupled map method is suitable to answer these questions, because the dynamics is decomposed into several elementary processes which are expressed by a simple dynamics (mapping). Hence, one can easily check the structural stability by replacing a procedure by another one.

In our model, the thermal diffusion and viscosity procedures are rather straightforward. Hence, we study the effects of modifying the buoyancy and pressure procedures by fixing the diffusion and viscosity ones. Although a variety of replacements can be considered, here, we restrict ourselves to the changes listed in table II.

Table II: Dynamics of the Procedures

| buoyancy dynamics | pressure dynamics |
|---|---|
| $v_y^* = v_y^t + c$ *discretized* $\partial^2 E/\partial x^2$ | $v_x' = v_x^* + \eta$ *discretized* $\nabla(\nabla \cdot \vec{v})$ |
| $v_y^* = v_y^t + c$ *discretized* $\partial^2 E/\partial y^2$ | $v_x' = v_x^* - \eta \exp(-\gamma v^2)$ |
| $v_y^* = v_y^t + c$ *discretized* $\partial^2 E/\partial x \partial y$ | $v_x' = v_x^* + \eta$ *discretized* $\nabla^4$ |

By choosing either one of the procedures listed in the table II, we have a total 9 possible models. Since, it is hard to report all simulations (onset of convection, routes to chaos, . . ., so on) for each model, we report mainly the onset of convection, spatiotemporal intermittency and the soft-hard turbulence transition.

At the onset of the Rayleigh-Bénard convection instability, we calculate the scaling property for the vertical velocity $v_y$ versus the normalized temperature difference $\epsilon$. We have found that the scaling property does not depend on the details of modeling. All the models that follow from the table II reproduce $v_y \sim \epsilon^{1/2}$, while the critical temperature difference $\Delta T_c$ depends on the models. This is reasonable since the scaling property is expected from the Hopf bifurcation analysis.

The distribution function of the laminar domains during STI depends on the choice of the dynamics, in particular, on the choice of the buoyancy procedure dynamics. For example, if we take a discretized $\partial^2/\partial y^2$ operator for buoyancy, the power law behavior is not obtained. In this case, laminar and globally turbulent states appear intermittently in time without spatial intermittency (i.e., spatial structure is not far from a homogeneous one).

The transition between soft and hard turbulences also depends on the change of the procedure. The replacement of the buoyancy procedure dose not affect the flatness of the temperature distribution. However, the pressure procedure is more relevant to the transition. When the pressure procedure is substituted by the cut off dynamics, the flatness rises from 2 to 10 with the change of $\Delta T$, (which does not depend on the cut off parameter $\gamma$). There is no plateau around 3 (corresponding to the Gaussian distribution; soft turbulence regime) and 6 (corresponding to the exponential distribution).

The results for the structural stability against the change of the procedure are summarized in the table III.

Table III: Reproducibility with changing Dynamics

| *Model* | $\partial_{xx} \cup \nabla(\nabla\vec{v})$ | $\partial_{yy} \cup \nabla(\nabla\vec{v})$ | $\partial_{xy} \cup \nabla(\nabla\vec{v})$ | $\partial_{xx} \cup \exp(\vec{v}^2)$ | $\partial_{xx} \cup \emptyset$ | $\partial_{xx} \cup \nabla^4$ |
|---|---|---|---|---|---|---|
| Onset | ◯ | ◯ | ◯ | ◯ | ◯ | ◯ |
| STI | ◯ | $+$ | $+$ | ◯ | ◯$^\dagger$ | ◯$^\dagger$ |
| SH | ◯ | ◯ | ◯ | $\times$ | $\times$ | $\times$ |

◯ : Agreement with experiments
$+$ : No spatial intermittency
$\dagger$ : Large fluctuation around the tail of the distribution $P(L)$
$\times$ : No Plateau around 3 (soft turbulence regime).

Generally, the choice of a procedure affects some property. The relevance of a physical process to a given behavior is determined by replacing the procedure corresponding to the process. For example, the scaling property near the onset of convection can be reproduced by a wide range of models, the pressure or buoyancy procedure is irrelevant to STI or the SH transition, respectively. The result that the buoyancy procedure does not play an important role for the SH transition implies that the mechanism of an external force does not affect the turbulence transition. In fact, the universal change of the distribution of a passive scalar is also observed in grid-generated turbulence and stirred fluids [51] [52].

To sum up, by determining the relevant procedures for a given behavior, it is possible to decide the universality class which yields the same salient behavior.

# B AutoRegressive Model

In the M-th order AR model, the time series $x(n)$ is expressed as

$$x(n) = \sum_{m=1}^{M} a_m x(n-m) + \epsilon(n) \tag{18}$$

where $\epsilon(n)$ is a residual error. We determine the coefficient $a_m$ by using the maximum likelihood method [18]. Of course we can obtain the power spectrum directly from the velocity time series. To get the frequency of the oscillation, however, we need a rather accurate form of the power spectrum, which requires a rather long computation. An advantage of the AR model is that the power spectrum inferred by it is a continuum function (rational polynomial function).

$$P(f) = \sigma^2 / ||1 - \sum_{m=1}^{M} a_m \exp(-2\pi i f n)||^2 \tag{19}$$

where

$$\sigma^2 =< \epsilon(n)\epsilon(n) > .$$

Then we can easily determine the characteristic frequency of the oscillation (see figure 34 which shows a power spectrum estimated by the AR model).

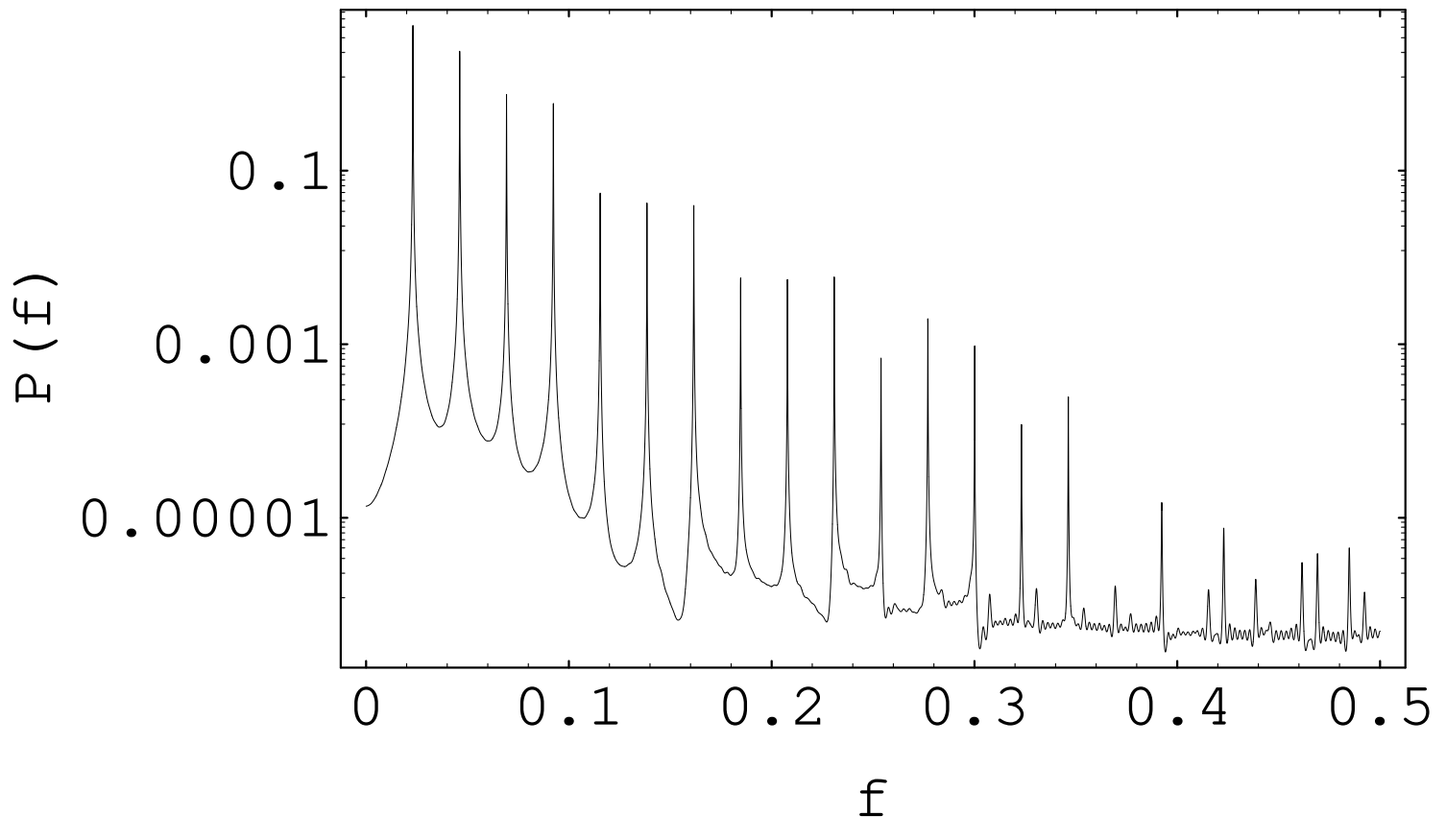

Figure 34: Power spectrum estimated by AR model

The power spectrum estimated by 100th order AR model near the onset of oscillation. Increasing $\Delta T$, the amplitude of oscillation gets larger and the higher harmonics starts to appear. The 4000 time series per 10 steps is used to determine the coefficient of AR model. $\Delta T = 0.3, \lambda = 0.4, \nu = \eta = 0.2, N_x = 17, N_y = 30$